%% file: book.tex
  \documentclass{cambridge6A}
  \usepackage{natbib}

  \usepackage{rotating}
  \usepackage{floatpag}
  \rotfloatpagestyle{empty}

  \usepackage{amsthm}
  \usepackage{graphicx}

  \usepackage{multind}
  \makeindex{authors}
  \makeindex{subject}

\usepackage{natbib}
\usepackage{longtable}

  \theoremstyle{plain}

  \theoremstyle{definition}

  \theoremstyle{remark}

  \usepackage{aas_macros}
  \usepackage[sectionbib]{chapterbib}

\begin{document}




  \pagenumbering{arabic}
 
  \mainmatter
  \include{Malzac/malzac}




\end{document}

%% file: Malzac/malzac.tex

\author[Julien Malzac]{JULIEN MALZAC}
\chapterauthor{JULIEN MALZAC}

\chapter[Radiation processes and models]{Radiation processes and models
\label{malzac}}

\contributor{Julien Malzac
\affiliation{University of Toulouse, France}}

\abstract{
\indent
This is a basic introduction to the physics of compact objects in the context
of High Time Resolution Astrophysics (HTRA). The main mechanisms of energy
release and  the properties of relevant radiation processes are briefly
reviewed. As a specific example, the top models for the multi-wavelength
variability of accreting black holes are unveiled.
}

\section{Introduction}

Compact objects represent a prime target for HTRA.  Their physics involves
strong gravitational fields, matter compressed to enormous densities, very high
energy particles and huge magnetic fields. Of course, such extreme conditions
cannot be produced in a laboratory on Earth. In many cases, high time resolution
observations of theses objects represent a unique opportunity to test
fundamental theories regarding particle interactions, the properties of dense
matter or gravitation.  Compact objects also inform us on fundamental
astrophysical processes such as accretion and ejection in their most extreme
form. 

Classically there are three types of compact objects that all  form mostly (but
not only) through the gravitational collapse of a normal star after exhaustion
of its thermonuclear fuel. White Dwarfs (hereafter WDs) have masses lower than
the Chandrasekhar limit  (1.4 $M_{\odot}$) and a radius comparable to that of
the Earth. The quantum degeneracy pressure of the electrons balances the
gravitation forces to prevent collapse. Neutron Stars (hereafter NSs) are
instead supported by short-range repulsive neutron-neutron interactions mediated
by the strong force and also by the quantum degeneracy pressure of neutrons.
They can have masses up to the Tolman-Oppenheimer-Volkoff (TOV) limit which is
approximately between 2--3\,$M_{\odot}$.  Above the TOV limit a star cannot
support its own weight.\footnote{Note that  objects supported by the quantum
degeneracy pressure of quarks, and named quark stars,  have also been theorised
and would have properties very similar to that of neutron stars, but their
existence is disputed.} It is completely collapsed and forms a Black Hole
(hereafter BH).  The size of a NS is about 10\,km, the size of a BH is given by
the size of the horizon of events.  The latter depends on the spin of the BH but
remains comparable to the gravitational radius: 
$R_{g}\,=\,GM_{\rm BH}/c^2\,\simeq\,1.5\,M_{BH}/M_{\odot}$\,km. 
A typical stellar mass BH concentrates about 10 times the
mass of the sun within a radius of about 15\,km. This corresponds to an average
density within $R_{g}$ of about $10^{15}$\,g\,cm$^{-3}$. The average density of
the matter of a NS is of the same order i.e. 10$^{14}$ times denser than planet
Earth!  The surface gravity on a NS is more than 10$^{11}$ times that on Earth. 
The dynamical time-scale in the environment of compact objects are very short.
For example the period of a Keplerian orbit at 6 times the radius of the object
corresponds, on Earth, to a geosynchronous orbit: it has a period of exactly 
one day. This time-scale is reduced to a few minutes around a WD and only a 
few milliseconds around a NS or a BH !  Compact objects can also release tremendous amounts of
energy in their environment in the form of radiation and powerful outflows. This
is this combination of power and very short dynamical time-scales makes compact
objects one of the main target of HTRA. Section\,\ref{sec:co} presents the basic
mechanisms of energy release around compact objects through dissipation of
gravitational, magnetic and rotational power. Then, in
Section\,\ref{sec:radproc}, the main emission processes  responsible for the
conversion of fraction of this energy into photons are described. Finally,
Sections\,\ref{sec:bhbaccretion} and \ref{sec:bhbjets} focus on models for the
multi-wavelength variability of accreting black hole in X-ray binary systems
involving respectively the accretion flow and the jets. 

\section{Compact objects}
\label{sec:co}

Energy can be extracted from the rotation of the compact object (Pulsars), its
magnetic field (magnetar), from the gravitational energy of material falling
onto the compact object (accretion in binary systems) or even Thermonuclear
fusion of outer layers of accreted materials (X-ray bursters, classical novae). 
In this section I  briefly introduce the different mechanisms of energy release.
A much more detailed exposition of compact objects physics can be found in
classical textbooks, such as those by 
\citet{1983bhwd.book.....S,1992apa..book.....F,1994hea2.book.....L,
2008bhad.book.....K}. 
There are also many excellent reviews covering specific recent developments  of
the field, for instance \citet{2012arXiv1206.2727P} on millisecond X-ray 
pulsars, \citet{2012ARA&A..50..609W} on thermonuclear burst oscillations,
\citet{2014PhyU...57..377G} on the structure of the boundary layer in accreting
neutron stars, \citep{2006ConPh..47..363S} on Cataclysmic variables, or
\citet{2007A&ARv..15....1D} on accreting BHs and NSs...  

 \subsection{Rotation}\label{sec:rot}
 
All stars are rotating (the rotation period of the sun is about 27\,d). In
the case of NSs and BHs the angular momentum is essentially conserved during the
gravitational collapse. The compactification of the star then implies faster
rotation.  As a result a newborn neutron star can have a spin period $P$ of the
order of $P$\,=\,10--100\,ms. On the other hand, most WDs have slow rotation due
removal of angular momentum during the loss of the progenitor's outer envelope
in a planetary nebulae. Nevertheless a WD can later  be spun-up by accretion
(see Section\,\ref{sec:accretion}). The fastest spinning WD has a period of only
$P\,=\,$13\,s \citep{2011ApJ...737...51M}.  The rotational energy of a star can be
estimated simply as $E_{\rm rot}\,=\,{I\Omega^2}/2\,=\,2I\pi^2P^{-2}$ where $I$  is the
moment of inertia which, for a homogeneous sphere of mass $M$ and radius $R$ is 
$I\,=\,2MR^2/5$. For a NS this energy can represent a small but significant fraction
of the rest mass energy of the star:

 \begin{equation}
 \frac{E_{\rm rot}}{M c^2}\simeq10^{-4} \left (\frac{R}{10 \, 
{\rm km}}\right)^2\left(\frac{P}{0.01 \,{\rm s}}\right)^{-2},
 \end{equation}

\noindent
 In absolute terms the rotation energy of a young neutron star is comparable to
the energy radiated by the sun during 1 billion years. However, this energy is
released over much smaller time-scales of the order of 10$^3$  to 10$^7$\,yr.
Indeed, the extraction of rotational energy implies spin-down. The rotational
power is related to the spin-down rate as 

 \begin{equation}
 \dot{E}_{\rm rot}\,=\,\frac{d}{dt}\left(\frac{I\Omega}{2}\right) \,=\,4I\pi^2\frac{\dot{P}}{P^3}\simeq 3\times 10^{50} \frac{M}{M_{\odot}} \left(\frac{R}{10 \, {\rm km}}\right)^2\left(\frac{P}{0.01 \,{\rm s}}\right)^{-2} \frac{\dot P}{P} \quad {\rm erg \, s^{-1}}.
 \end{equation}

\noindent
The `braking' time-scale over which the rotation power is released can be
estimated as $E_{\rm rot}/{\dot E_{\rm rot}\sim P/\dot P}$. Pulsars are NSs
emitting beams of radiation leading to the highly coherent modulation of the
observed light curve at the spin period. In these sources both the spin and spin
down rate can be measured accurately (see Chapter\,5 by A.\,Possenti in these 
proceedings)
and this allows one to estimate the released rotation power. In the case of the
Crab pulsar the observed period is $P\,\simeq\,33$\,ms and spin-down rate is
$\dot{P}/P\,\simeq\,10^{-11}$\,s$^{-1}$ which leads to $\dot{E}_{\rm rot}\,\simeq\,
10^{38}$ erg s$^{-1}$. This power is comparable to the accumulated radiative
output of about $10^{5}$ stars like the Sun. This power is larger than the
observed luminosity of the source but comparable to the power required to feed
the surrounding pulsar wind nebula. This coincidence indicates that the pulsar
wind is powered by the rotational energy of the pulsar. 

\noindent
The rotational power is extracted via the effects of the strong magnetic 
field of the neutron star. Indeed, during the gravitational collapse 
leading to the formation of the NS, the conservation of magnetic flux 
across the stellar surface implies that the magnetic filed is amplified 
by a factor $\sim\,R_{\star}^2/R^2 $, where $R_{\star}$ is the initial 
radius of the star and $R$ is the radius of the compact object. For a 
star of the size of the sun compacted into a 10 km radius, this 
amplification factor is of order of $5\times 10^9$ !  As a result the 
typical surface magnetic field of a young neutron star is of the order of 
10$^{11}$-10$^{12}$\,G. For comparison, the Earth magnetic field is $\sim\,0.5$\,G, a fridge magnet is 
$\sim\,50$\,G, the strongest continuous field 
yet produced in a laboratory is about 5$\times 10^{5}$\,G. Because of the 
high conductivity of the NS matter, its magnetic field dissipates only 
very slowly, on times scales much longer than the `braking' time-scale of 
the NS. In practice it can be considered a constant.  The young NS 
therefore behaves has a huge spinning magnet. It is known from 
electromagnetic theory that a magnetic dipole rotating in vacuum 
radiates. The radiated energy can be estimated simply as

 \begin{equation}
 \dot{E}_{d}\,=\,\frac{B_{p}^2R^6\Omega^4}{6c^3}\sin^2{\alpha}\simeq 10^{39}\left(\frac{B_p}{10^{12}\,{\rm G}}\right)^2\left(\frac{R}{10\,{\rm km}}\right)^6\left(\frac{P}{0.01 \, {\rm s}}\right)^{-4} \sin^2{\alpha} \quad {\rm erg s^{-1}},
 \end{equation}

\noindent
where $B_{p}$ is the magnetic field amplitude at the poles of the compact star
and $\alpha$ is the misalignment angle between the direction of the dipole and
the spin of the neutron star. This radiated power can be matched  to the
observed rotational power in pulsars in order to estimate the pulsar magnetic
field, its age etc... But in reality the magnet is not spinning in vacuum. Huge
electric fields are generated close to the surface of the NS. Those fields
extract charged particles from the NS. These charged particles distribute
themselves around the star to neutralise the electric field: an extended
magnetosphere is formed. Streams of charged particles leave the star at high
latitudes where the magnetic field lines are open, leading to the formation of a
wind. The rotation energy is dissipated mostly through the interaction of the
magnetic field of the NS with the magnetosphere and the wind. Nevertheless the
energy losses remain comparable to that predicted by the simple magnetic dipole
radiation model. 
 
\subsection{Magnetic field dissipation}\label{sec:mag}
  
Some NSs appear to have negligible rotation energy but huge magnetic 
fields up to 10$^{14}$-10$^{16}$\,G. These strongly magnetised NSs define 
the class of Magnetars \citep[see e.g.][]{2006csxs.book..547W}.  It has been 
shown \citep{1993ApJ...408..194T} that such magnetic fields may have built 
up through dynamo processes (similar to that generating the Earth or the 
sun magnetic fields) during the first moments of the life of the NS. This 
strong dynamo amplification occurs only if the NS is formed with a spin 
period $P<5$\,ms. After 10--30\,s, the core has cooled down and is not hot 
enough for efficient dynamo. The magnetic braking is very efficient and 
the rotation energy is dissipated very quickly. After few minutes the 
rotation has slowed down to a period of a few seconds. But the magnetic 
fields remain so strong that they can push and move material around in 
the star interior and crust. This leads to large amounts of magnetic 
dissipation during the first 10$^4$\,y. Magnetic dissipation in the 
interior of the star keeps the star hot and bright producing mostly X-ray 
thermal emission. Magnetic dissipation may also occurs in the surrounding 
magnetosphere due to twisting of the magnetic field lines and 
reconnection (similar to solar corona). This can lead to bursts of 
non-thermal X-ray and gamma-ray radiation which are observed in sources 
called `soft gamma-ray repeaters'.

 \subsection{Accretion}
\label{sec:accretion}

Accretion is the growth of a massive object by gravitationally attracting 
more matter. This is an ubiquitous astrophysical process leading to the 
formation of planets, stars, galaxies and the growth of super-massive BHs. 
The gravitational energy released during accretion onto super-massive BHs 
also appears to regulate the joint growth of the BHs and their host 
galaxy. Accretion onto stellar compact objects can be observed if the 
compact objects is in a binary system and can accrete gas from its 
companion star. As this type of accretion occurs in bright nearby sources 
evolving on human time-scales it is relatively easy to observe and study, 
and some of the knowledge gained may be extrapolated to other accreting 
systems such as super-massive BHs in Active Galactic Nuclei (AGN). As will 
be discussed below accretion onto a compact star in an accreting binary 
system depends on the nature of the donor star (high mass vs low mass 
star). It depends also on the nature of the compact star. Gas accreting 
onto a NS or a WD will ultimately hit the hard surface of the star. This 
has observable effects such as the presence of a boundary layer in which 
the gas is stopped, or the triggering of nuclear explosions (X-ray bursts 
in NS, classical novae in WD) when enough material has been accreted onto 
the surface. The structure of the accretion flow may also be affected by 
the strong magnetic field of the star as is the case for NS in X-ray 
pulsars or WD in polars. Such complications do not occur in accreting BHs 
where we can observe accretion in its `purest' form: as the gas crosses 
the event horizon without notable effects, all the observed radiation 
must originate from the accretion flow.

\subsubsection{Accretion power}

Accreting matter falls into the potential wells of a compact object and 
loses gravitational energy.  For accretion to occur gravitational energy 
(and angular momentum) must be dissipated away, mostly in form of 
radiation.  If accretion occurs at a mass accretion rate $\dot{M}$ onto 
an object of size $R$ and mass $M$ the gravitational power that is 
dissipated is given by the gravitational potential on the surface

\begin{equation}
\dot{E}_{\rm ac}\,=\,\dot{M}\frac{GM}{R}\,=\,\eta \dot{M}c^2,
\end {equation}

\noindent
where $\eta\,=\,GM/Rc^2$ is called the accretion efficiency. It represents 
the amount of energy released per unit of mass energy accreted. The 
accretion efficiency onto a WD is of the order of $10^{-4}$ while 
accretion onto a neutron star reaches 0.1. Due to the absence of a hard 
surface when accreting onto a BH, the efficiency depends on the structure 
of the accretion flow. For thin accretion discs (see below) the accretion 
efficiency is in the range 0.057--0.42 depending on the BH spin. For 
comparison accretion onto planet Earth has $\eta\,\sim\,10^{-9}$, while the 
Sun has $\eta\,\sim\,10^{-6}$. The efficiency of thermonuclear fusion of 
hydrogen is 7$\times 10^{-3}$. Accretion onto a compact object is 
therefore much more efficient than fusion at releasing energy.

\subsubsection{Eddington limit}

The accretion power is however limited by the amount of matter that we 
are able to accrete. There is a fundamental limit on the mass accretion 
rate that is set by the Eddington luminosity. The Eddington luminosity is 
the maximum luminosity for which the gravitational force on a fluid 
element exceeds the radiation pressure (i.e. the maximum luminosity at 
which matter can be accreted). Let us consider a fluid element of mass 
$m$ located at a distance $d$ from the center of the compact object which 
radiates isotropically a luminosity $L$. The amplitude of the force of 
gravity is $F_{\rm grav}\,=\,GMm/d^2$. The radiation pressure force is 
proportional to the local radiation flux and is directed in opposite 
direction. It is given by $F_{\rm rad}\,=\, L\kappa m/(4\pi\,d^2 c)$, where 
the opacity $\kappa$ is a measure of the effectiveness of the transfer of 
momentum from radiation to the fluid element. Since both forces are 
proportional to $m/d^2$. The condition for equilibrium $F_{\rm 
grav}\,=\,F_{\rm rad}$ is independent of both the mass and distance of the 
fluid element. This condition defines the Eddington luminosity

\begin{equation}
 L_{\rm E}\,=\,\frac{4\pi G M c}{\kappa}\simeq 1.46 \times 10^{38} \frac{M}{M_{\odot}} \quad {\rm erg s}^{-1}.
 \label{eq:le}
\end{equation}

\noindent
The accreting gas is usually hot and ionised, so that the opacity is dominated by
electron scattering. In this case $\kappa\,=\,\kappa_{es}\,=\,0.34$ cm$^2$ g$^{-1}$ for standard
abundances. This is the value of $\kappa$ that  was assumed in the  numerical estimate
given in equation\,\ref{eq:le}. This immediately implies a maximum accretion rate

\begin{equation}
\dot{M}_{\rm E}\,=\,L_{\rm E}/\eta c^2 \,=\, 2.6 \times 10^{-9} \left(M/M_{\odot}\right) \eta^{-1} \quad M_{\odot}\, {\rm yr}^{-1}
\end{equation}

\noindent
For a compact object with a hard surface (NS or WD), the Eddington mass 
accretion rate depends only on its size

\begin{equation}
\dot{M}_{\rm E}\,=\,\frac{4\pi c R}{\kappa} \,=\,1.8 \times 10^{-8} \left(\frac{R}{10 \, {\rm km}}\right)\quad  M_{\odot} \, {\rm yr}^{-1} 
\end{equation}

\noindent
This maximum luminosity and accretion rate are estimated for a spherical 
accretion geometry and radiation field, and deviation from spherical are 
of course expected to occur and may allow to exceed somewhat the 
Eddington limit (which indeed appears to be violated in some sources). 
Nevertheless this gives a good estimate of the maximum power that can be 
extracted through accretion onto a compact object. This power is huge. 
The Eddington luminosity of a 10\,$M_{\odot}$ BH is comparable to the 
combined luminosity of 10$^6$ stars like the Sun.

\subsubsection{Mass transfer} 

However there is another limitation related to the capacity of the 
compact object to attract and capture gas from the donor star. This is 
the so-called mass transfer problem. Before falling onto the compact 
object the gas must escape the pull of the donor star.  The effective 
gravitational potential in a binary system is determined by the masses of 
the stars and the centrifugal force arising from the motion of the two 
stars around one another. One may write this potential as a function of 
$r_1$ and $r_2$ the distance to the center of each star and $r_3$ the 
distance to the rotation axis

\begin{equation}
\phi\,=\,-\frac{GM_1}{r_1}-\frac{GM_2}{r_2}-\frac{\Omega_{\rm orb}^2 r_3^2}{2},
\end{equation}

\noindent
where $\Omega_{\rm orb}$ is the orbital angular velocity. The 
equipotential surfaces form two lobes surrounding each of the stars that 
are called the Roche lobes. The two lobes are connected through a point 
called the first Lagrangian point (or L1) where the sum of the 
centrifugal and gravitational forces vanishes. This is a saddle point in 
the potential which forms a pass that the gas from the donor has to climb 
before being able to fall into the influence of the compact object.

\paragraph{Roche lobe overflow}

Mass transfer may occur simply if the donor fills its Roche lobe. This 
may result from an increase of the stellar radius during the evolution of 
the star. This can also be driven by changes in the orbital parameters 
that make the Roche lobe smaller. This can also occur occur through loss 
of angular momentum (by emission of gravitational waves, magnetic 
braking, tidal torques, mass loss in a stellar wind...). Also mass 
transfer, over time, will change the mass ratio of the two components and 
affect the orbital parameters. So even if the donor fills its Roche lobe, 
accretion may be instable and may not be sustained. It can be shown that 
stable lobe overflow can occur only if the mass of the donor is smaller 
than the mass of the accretor. If these two conditions are fulfilled then 
steady mass accretion occurs at rates of the order of $\dot{M}\,\sim\,10^{-10}$--$10^{ -9}$. Such systems are called low-mass X-ray binaries 
(hereafter LMXB).

\paragraph{Wind accretion}

A massive early type companion (O or B) can loose mass in a wind at a 
rate 10$^{-6}$--10$^{-5}$ $M_{\odot}$ yr$^{-1}$ and supersonic velocity 
comparable to the escape velocity of the star

\begin{equation}
v_{\rm w}\sim v_{\rm esc}\,=\,\sqrt{\frac{2GM_{\star}}{R_{\star}}} \sim 10^3 \quad{\rm km \, s}^{-1}
\end{equation}

\noindent
The compact star will gravitationally capture matter from a roughly cylindrical region
with axis along the relative wind direction. This cylinder represents the volume where
the wind particle kinetic energy is less than the gravitational potential 
\citep[e.g.][]{1944MNRAS.104..273B, 1952MNRAS.112..195B, 1973ApJ...179..585D, 
1973ApJ...184..271L, 1992apa..book.....F}.
The radius  of the cylinder, called the accretion radius or
gravitational capture radius, is given by 

 \begin{equation}
r_{\rm acc}\simeq \frac{2GM}{v_{\rm rel}^2+c_{\rm s}^2}
\end{equation}

\noindent
where $c_{\rm s}$ is the sound speed in wind, $v_{\rm rel}^2\,=\, v_{\rm orb}^2+v_{\rm
w}^2$, and $v_{orb}$ is the orbital velocity of the gas.

\noindent
The net amount of gas captured and accreted by the compact object can be 
obtained by combining this relationship with Kepler's third law and the 
continuity equation (assuming spherically symmetric and steady mass 
loss)

 \begin{equation}
 {\dot{M}}\,=\,\pi r_{\rm acc}^2\rho v_{\rm rel}\sim (10^{-5}-10^{-4})\dot{M}_{\rm w}\sim 10^{-11}-10^{-9}  M_{\odot} \,{\rm yr}^{-1}
 \end{equation}

\noindent
We can see that both wind accretion in HMXB an Roche lobe overflow in 
LMXB allow mass transfer at a rate that is smaller than, and yet a 
significant fraction of, the Eddington limit.

\subsubsection{Keplerian accretion discs}\label{sec:kepad}

Once the material is captured, it would orbit the compact object 
indefinitely unless it can get rid of its angular momentum and be 
accreted. It is viscosity, and the associated viscous torques between 
annuli in the disc, that allows angular momentum to be transferred 
outwards and mass to spiral inwards. Any original ring of particles will 
thus spread out into a disc, with the outer radius being determined by 
tidal torques from the companion. However, ordinary molecular viscosity 
is completely inadequate to account for the observed properties of discs 
and some kind of turbulent viscosity is invoked.  The best candidate as 
the origin of this turbulent viscosity is the Magneto-rotational 
instability which develops in differentially rotating flows and can 
generate fully developed Magneto-HydroDynamic (MHD) turbulence that 
provides efficient angular momentum transport 
\citep{1991ApJ...376..214B,1998RvMP...70....1B}. The characteristic scale 
of the turbulence must be less than the disc thickness, $H$, and the 
characteristic turbulent speed is expected to be less than the speed of 
sound, $c_{\rm s}$, since there is no strong evidence for turbulent 
shocks. The viscosity, $\nu$, is therefore often parametrised by writing 
it as $\nu \,=\, \alpha c_{\rm s} H$, placing all the uncertainties in the 
unknown parameter $\alpha$, taken to be less than 1. This is the standard 
alpha-disc model of accretion discs \citep{1973A&A....24..337S}. This 
model also assumes that radiation cooling in the disc is very efficient. 
All the locally dissipated accretion power is radiated away on the spot 
in the form of blackbody radiation. As a consequence the disc is cold and 
geometrically thin. From these two assumptions, it is possible to 
determine analytically the full disc structure as a function of the 
distance $r$ from the compact object. Although there is still no reliable 
way of calculating $\alpha$, it turns out that the observable properties 
of a steady accretion disc are largely independent of $\alpha$. In 
particular the total luminosity of the disc is simply half of the 
gravitational power at the disc inner radius $R_{i}$, $L_{\rm disc}\,=\, 
2GM\dot{M}/R_{i}$. Most of the luminosity comes from the inner part of 
the accretion flow, 80\,\% of the power is radiated within 10$R_i$.  The 
temperature profile in the disc is

\begin{equation}
T(R)\,=\,T_{0} \left(\frac{R_{i}}{r}\right)^{3/4}\left(1-\sqrt{\frac{R_{i}}{r}}\right)^{1/4},
\end{equation}

\noindent
where

\begin{equation}
T_{0}\,=\,\left(\frac{3GM\dot{M}}{8\pi\sigma R_{i}^3}\right)^{1/4}\simeq 6\times 10^7 \left(\frac{R_i}{R_{G}}\right)^{-3/4}\left(\frac{\dot{M}c^2}{L_{E}}\right)^{1/4}\left(\frac{M}{M_{\odot}}\right)^{-1/4}     \quad {\rm K}.
\end{equation}

\noindent
The temperature has a maximum around 3$R_i$ and then decreases with the 
distance like $r^{-3/4}$. This maximum temperature can reach a few keVs 
in NSs and BHs. The innermost and most luminous part of the accretion 
disc will therefore produce mostly X-ray radiation. The spectral energy 
distribution (SED) of the whole accretion disc is constituted of the sum 
of all the blackbodies emitted at different radii in the disc with a 
different temperatures. Longer wavelengths allow one to probe more 
distant regions of the accretion flow. We note the standard 
Shakura-Sunyaev disc model assumes pure Newtonian physics, the general 
relativistic version of the thin disk model was introduced by 
\citet{1973blho.conf..343N}.

\subsubsection{Hot accretion flows}

The thin accretion disc model assumes  full energy thermalisation and this
implies a low temperature, a high density and a small disc scale height
$H/R\,\sim\,10^{-3}$--10$^{-4}$. If instead, the disc is hot, then the  gas pressure
makes the scale height larger $\simeq$\,0.1--0.5 and this reduces the density. As
a consequence the radiation cooling is less efficient and a high temperature can
be maintained.  Indeed, the density becomes so small that the collision
time-scales between electrons an ions of the plasma can be long compared to the
accretion time-scale. Then the protons acquire most of the gravitational energy
through viscous heating  but this energy can only be radiated efficiently by the
electrons. Since electrons and protons  are decoupled, the two populations end
up having very different temperatures. The protons can reach $\sim\,10^{12}$\,K
while the electrons are much colder (and yet very hot), $\sim\,10^{9}$\,K  in the
innermost part of the accretion flow. In this kind of hot accretion flow the
accretion energy is not radiated away locally, it can be carried inward with the
flow until it is swallowed by the BH or hit the surface of the compact star.
These advection dominated accretion flows (Ichimaru 1977; Narayan \& Yi 1994)
produce mostly non-thermal  Comptonised radiation (see Section\,\ref{sec:thcomp}).
Hot accretion flows have been extensively studied. Analytic solutions have been
found taking also into account the effects of convection; 
convection dominated accretion flows \citep[CDAF,][]{2000ApJ...539..798N} 
and outflows; advection dominated inflow-outflow solution)
\citep[ADIOS,][]{1999MNRAS.303L...1B}. Hot accretion flow have
also been found in numerical simulations  \citep{1999MNRAS.310.1002S}. 

\subsubsection{Jet launching from accretion flows}

Accreting NS and BH (an perhaps also WD) can launch highly collimated 
jets of magnetised plasma that travel at near light speed, carrying away 
a significant fraction of the accretion power \citep{1994Natur.371...46M, 
2014SSRv..183..323F}. These jets propagate over large distances and can 
have an enormous impact on the surrounding medium over distance scales 
that are far out of the gravitational reach of the BH itself. The details 
of the formation of those jets are unclear, none of the models and 
simulations take into account all the physics. But all models require 
magnetic fields. The mechanism proposed by \citet{1982MNRAS.199..883B} 
involves magnetic field lines threading the accretion disc. Provided that 
the magnetic field lines are are sufficiently inclined with respect to 
the disc, the centrifugal force can throw away some of the accreting 
material, which remains tied to the rotating magnetic field lines like a 
bead on a wire. Then the jet needs to somewhat collimate and this 
requires large scale coherent magnetic field structures. Another 
mechanism involves electromagnetic extraction of energy from a Kerr BH. 
Indeed, \citet{1969NCimR...1..252P} and \citet{1970PhRvL..25.1596C} have 
shown that up to 30\,\% of the mass energy of a maximally spinning BH 
can be theoretically extracted. Then, \citet{1977MNRAS.179..433B} have 
shown that an accretion disc can allow this energy to be extracted and 
drive powerful jets. The magnetic field carried by the accreted gas 
remain threaded through the BH horizon.  The frame of the field lines is 
dragged along with the rotation of the BH. These rotating field lines 
induce an electromagnetic force that accelerates charged plasma at 
relativistic speeds along the axis of rotation. Due to the radial 
component of the field, the particles spiral as they leave.

\subsubsection{Effects of the  magnetic field  of a compact star on the accretion flow}

Magnetic fields become dynamically important close to the compact object 
at a distance called the Alfven radius, $R_{\rm a}$. At $R_{\rm a}$ the 
magnetic energy is comparable to the kinetic energy of the infalling gas: 
$\rho v^2/2\,\simeq\,B^2/8\pi$. For the purpose of simple estimates, we can 
assume that the accreting gas is in free fall in a spherical geometry, so 
that $v\,\simeq\,v_{\rm ff}\,=\,\sqrt{2GM/R_{\rm a}}$ and $\rho\,=\,\dot{M}/(4\pi 
R_{\rm a}^2v_{\rm ff})$. Assuming a dipole magnetic field around a WD or 
NS of radius $R_{\star}$: $B(R_{\rm a})\,\sim\,B_p\,R_{\star}^3/R_{a}^3$, we 
then obtain

\begin{equation}
R_{\rm a}\simeq 30 \, {\rm km} \left(\frac{B_{p}}{10^9 \, G}\right)^{\frac{4}{7}}\left(\frac{\dot{M}}{2\times 10^{-8} \,M_{\odot} \,{\rm yr}^{-1}}\right)^{-\frac{2}{7}}\left(\frac{M}{M_{\odot}}\right)^{-\frac{1}{7}}\left(\frac{R_{\star}}{10 \, {\rm km}}\right)^{\frac{12}{7}}.
\end{equation}

\noindent
The effects of the magnetic field are important only if $R_{a}>R_{\star}$, 
which translates into the following condition for the surface magnetic field:

\begin{equation}
B_p > 1.5 \times 10^8  \left(\frac{\dot{M}}{2\times 10^{-8} \,M_{\odot} \,{\rm yr}^{-1}}\right)^{\frac{1}{2}}\left(\frac{M}{M_{\odot}}\right)^{\frac{1}{4}}\left(\frac{R_{\star}}{10 \, {\rm km}}\right)^{-\frac{5}{4}} \quad {\rm G}.
\label{eq:bpmin}
\end{equation}

\noindent
As discussed above in Sections\,\ref{sec:rot} and \ref{sec:mag}, in the 
case of NS this condition will be easily verified and the magnetic fields 
of the star will interfere with the accretion process.  At $R_{\rm a}$ 
the magnetic field can force accreting material in co-rotation with the 
compact star.  If the spin period of the compact star is longer than the 
orbital period at $R_{\rm a}$, the centrifugal forces cannot balance 
gravity anymore and the material flows along magnetic field lines onto 
the magnetic poles of the compact star. This forms an X-ray pulsar if the 
compact star is a NS, and a polar in the case of a WD.  The magnetic 
field transfers angular momentum from the accretion flow to the compact 
star exerting an effective spin up torque on the star. If the spin period 
of the compact star is shorter than the orbital period at $R_{\rm a}$, 
the corotation implies that the centrifugal forces overcome gravity: 
accretion is stopped. This is the so called propeller regime, in which 
angular momentum is transported from the compact star to the `accretion' 
flow.  In this case the magnetic field exerts a spin-down torque on the 
star.  The spin of the compact star therefore evolves through propeller 
and accreting regimes to reach an equilibrium in which the spin period of 
the star is equals to the orbital period at $R_{\rm a}$

\begin{equation}
P_{\rm eq}\simeq 3 \,{\rm ms }\left(\frac{B_{p}}{10^9 \, G}\right)^{\frac{6}{7}}\left(\frac{\dot{M}}{2\times 10^{-8} \,M_{\odot} \,{\rm yr}^{-1}}\right)^{-\frac{3}{7}}\left(\frac{M}{M_{\odot}}\right)^{-\frac{5}{7}}\left(\frac{R_{\star}}{10 \, {\rm km}}\right)^{\frac{18}{7}}.
\end{equation}

\noindent
This equation shows that accretion will spin up a NS up to spins of a few ms.
Such sources are observed, these are the millisecond X-ray pulsars 
\citep{1998Natur.394..344W, 2013FrPhy...8..679H, 2015arXiv150103882T}.

\subsubsection{Compact stars with weak magnetic field: boundary layer}

Many compact stars appear to have a magnetic field below the threshold 
defined by equation\,\ref{eq:bpmin}. Indeed, accretion seems to cause 
dissipation of the magnetic field which may be considerably reduced in 
the course of the history of the accreting compact star.  In this case 
the accretion disc can extend undisturbed very close to the surface of 
the star. There is however a region between the accretion flow and the 
star where the accreting gas has to decelerate from the orbital velocity 
to the rotation velocity of the star, spinning up the compact star in the 
process. This region is called the boundary layer (BL).  Since most of 
the kinetic energy of the gas is dissipated in the BL the luminosity of 
the BL is comparable to the total luminosity of the accretion disc: 
$L_{\rm BL}\,\sim\,\dot{M}\,v_{\rm K}^2/2\,=\,GM\dot{M}/(2R_{\star})\,\sim\, L_{\rm 
disc}$. Matter has a significant latitude velocity component in the 
boundary layer, spreading above the compact star surface and decelerating 
due to friction (like a wind above the sea, \citet{1999AstL...25..269I}. 
For sources with luminosity greater than 5\,\% Eddington, the 
local radiation flux is Eddington. In this regime the spreading layer 
temperature is independent of luminosity. Emission models predicts a 
temperature of the order of 2.5\,keV which was confirmed by observations. 
The size of the belt must increase with accretion rate/luminosity. For 
$L\,\sim\,L_{\rm E}$, the spreading layer covers the whole surface of the 
compact star \citep{2014PhyU...57..377G}.

\section{Radiation processes}\label{sec:radproc}

From radio waves to gamma-rays, compact objects radiate over the whole 
electromagnetic spectrum. This radiation can be emitted through thermal 
radiation (i.e. blackbody emission related to the temperature of the 
object) or non-thermal processes like Synchrotron or inverse Compton. 
Synchrotron radiation is produced by very energetic charged particles 
spiralling around magnetic field lines. Relativistic particles 
accelerated in shocks in the jets produce synchrotron emission mostly in 
the radio, sub-mm and infrared (IR) bands.  A similar process, called 
curvature radiation, is related to the propagation of particles along 
very curved field lines. It can be important in the magnetosphere of 
neutron stars. An accretion flow can be very hot and contains energetic 
electrons of temperature up to a billion K. It also contains many low 
energy photons produced by synchrotron radiation or coming from the outer 
regions of the accretion flow. These photons collide with the hot 
electrons and gain energy at each interaction.  This process is called 
Inverse Compton scattering. If the photon can make several interactions 
before escaping the hot gas, the process is called Comptonisation. It 
leads to the production of hard X-ray radiation. Inverse Compton from 
relativistic particles in the jets may also produce gamma-rays.  
Understanding the radiation processes allows one to extract the 
information that is encoded in the radiation received from these objects. 
It enables to measure many physical parameters of the system such as 
temperatures, velocities, magnetic fields, the energy of accelerated 
particles and their distribution. It also helps to determine the size and 
geometry of the emitting regions and test the predictions of theoretical 
models.  By simultaneously observing in different bands we learn about 
the joint evolution of the different components of the systems (e.g. 
accretion flow and jets).  In this section, I summarize the main 
properties of the cyclo-synchrotron, curvature radiation, Bremsstrahlung, 
inverse Compton and photon-photon pair production. A more detailed 
discussion of these radiation processes can be found in classical text 
books such as those by \citep{1986rpa..book.....R, 1992hea..book.....L, 
Jackson99}.

\begin{figure} 
\includegraphics[scale=0.30]{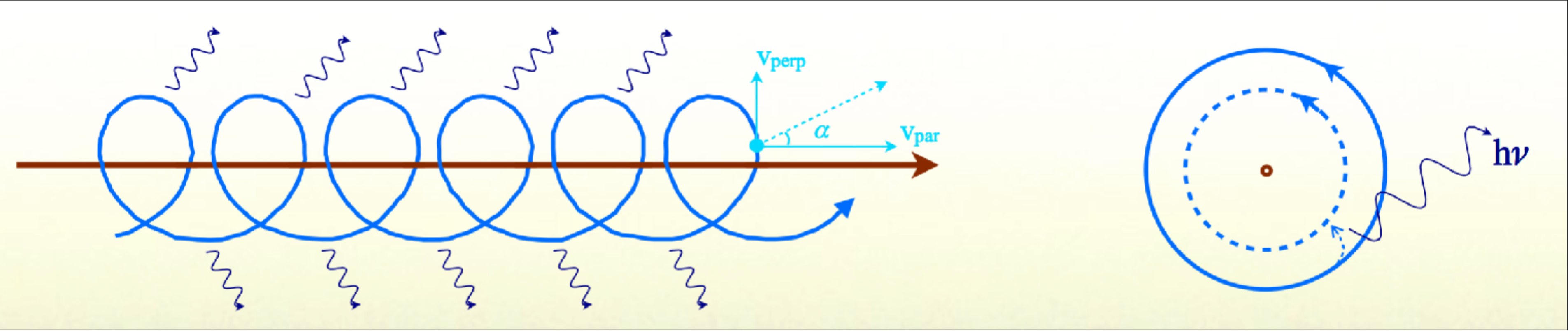} 
\caption{Emission of an electron spiralling in a magnetic field} 
\label{fig:synch} 
\end{figure}

\subsection{Cyclo-Synchrotron}

Cyclo-synchrotron radiation is the radiation produced by a charged 
particle accelerated by the magnetic field.  In the following I consider 
an $e^{-}$ which gyrates around the magnetic field lines with a velocity 
$v$ (see Figure\,\ref{fig:synch}). The component of its velocity which is 
parallel to the magnetic field is a constant. The transverse component of 
the velocity has a constant modulus and a direction that is rotating 
uniformly around the magnetic filed line. The constant angle $\alpha$ 
between the velocity and the magnetic field is called the pitch angle.  
The rotation frequency $\nu_{B}\,=\,\nu_{\rm L}/\gamma$, where 
$\gamma\,=\,\left(1-v^2/c^2\right)^{-1/2}$ is the Lorentz factor of the 
particle and $\nu_{\rm L}\,=\,qB / 2 \pi m_e c\,=\, 2.8 \times 10^6 \, {\rm 
B}$\,Hz, is the Larmor frequency. The magnetic field $B$ is expressed in 
$G$. The radius of the `orbit' around the magnetic field is given by the 
Larmor radius: $r_{L}\,=\,p_{\perp} c/ (2 \pi \nu_{\rm L})\,=\,1.7 \times 10^3 
p_{\perp}/B$ cm, where $p_{\perp}\,=\,p \sin{\alpha}\,=\,\gamma v \sin{\alpha} 
/c$. Because the particle has a transverse acceleration it radiates. The 
cyclo-synchrotron emission is calculated under a number of assumptions. 
First, the magnetic field must be uniform at the Larmor scale. This 
implies for instance that there is no turbulence at that scale. The 
magnetic field lines must not be too curved as this can be the case in 
the magnetosphere of pulsars for example.  Moreover, as the particle 
radiates, its kinetic energy and Larmor radius both decrease. The energy 
loss must be small on the a giration time-scale, in other words we 
require that the cooling time-scale the particle $t_{\rm cool} 
>>1/\nu_{\rm B}$.  Finally we will present the results obtained in the 
classical limit in which the magnetic field is smaller than the critical 
limit $B_{\rm c}\,=\,m_e c^3/\hbar q\,=\,4.4. \times 10^{13}$\,G. Above this limit 
the particle energies and orbits around the magnetic field become 
quantised.

\subsubsection{Power} 

The radiated power of an accelerated particle is given by the Larmor formula 

\begin{equation}
P\,=\,\frac{2q^2}{3c^3}\gamma^4\left(a^2_{\perp}+\gamma^2a_{\parallel}\right),
\label{eq:larmorform}
\end{equation}

\noindent
where $a_{\perp}$ and $a_{\parallel}$ represent the component of the 
acceleration that are perpendicular and parallel to the velocity 
respectively.  In the case of particle girating in the magnetic field 
$a_{\parallel}\,=\,0$ and $a_{\perp}\,=\,\nu_{\rm B}\,v_{\perp}/2\pi$. The 
radiated power can be written as 
$P\,=\,2c\sigma_{\rm T}\,U_{\rm B}\,p^2_{\perp}$, where $U_{\rm B}\,=\,B^2/8\pi$ is the magnetic field energy 
density, and $\sigma_{\rm T}$ is the Thomson cross section (see 
Section\,\ref{sec:compton}). Usually the distribution of pitch angles is 
isotropic. This is either because the distribution of radiating particles 
is isotropic, or the magnetic field is isotropically tangled, or both. 
Then the average over the pitch angles gives

\begin{equation}
P\,=\,\frac{4}{3}c\sigma_T U_{\rm B} p^2.
\label{eq:synchpow}
\end{equation}

\noindent
Therefore the requirement that the cooling time of a particle $t_{\rm
cool}\,\simeq\,\gamma m_{\rm e}c^2/P$ is longer than the giration time-scale
$1/\nu_{\rm B}$, implies a limit on the product of the particle energy and
magnetic field

 \begin{equation}
 \gamma^2 B< 2q/r_0^2.
 \label{eq:cond} 
\end{equation}

\subsubsection{Optically thin synchrotron spectrum}

If the charged particle moves very slowly around the magnetic field 
lines, the amplitude of the electric field measured at large distance is 
proportional to the sine of the angle between the acceleration vector and 
the direction to the observer.  This angle changes with time $t$ as the 
particle `orbits' the magnetic field and the observer observes a 
sinusoidal modulation of the electric field: $E(t)\,=\,\sin{2\pi\nu_{\rm 
B}t}$.  The emitted spectrum is proportional to the Fourier power 
spectrum of the fluctuations of the electric field which, in this case is 
a Dirac delta function. In other words, the observer sees a line at 
frequency $\nu_{\rm B}$. If the particle is moving faster, the emission 
pattern is modified by Doppler beaming and harmonics appear in the 
spectrum. For a relativistic particle, the emission is beamed within an 
angle $1/\gamma$ around the direction of the velocity.  When this 
emission cone crosses the direction to the observer, she measures a 
briefly increased electric field. The duration of this pulse is $\delta t 
\,\simeq \,1/(\nu_{\rm B} \gamma^3)$\footnote{The line of sight remains in 
the emission cone during $1/(\nu_{\rm B} \gamma)$, the additional 
$\gamma^{-2}$ factor is caused by photon travel time effects.}. This 
pulsed modulation of the electric field at frequency $\nu_{B}$ is made of 
multiple lines at frequencies that are multiple of $\nu_{B}$ and 
extending up to a frequency $\nu_{\rm c}$ that is comparable to the time 
scale of the pulse:

 \begin{equation}
 \nu_{\rm c}\,=\,\frac{3}{2}\gamma^3\nu_{B} \sin{\alpha}.
\end{equation}

\noindent
In the ultra-relativistic case ($\gamma >>1$), $\nu_{B}$ vanishes and the 
spectrum tends toward a continuum. This limit is called Synchrotron 
emission, while the non-relativistic case is called cyclotron emission.  
In fact once one consider not only the emission from a unique particle 
but that of a set of particles with different directions of propagation 
and pitch angles the harmonic lines are considerably broadened, even in 
the mildly-relativistic case (see \citealt{2003A&A...409....9M} for a 
general calculation of this case).

Equation.\,\ref{eq:synchpow} shows that the emission increases quadratically with the momentum
of the particle, the radiation induced by
the magnetic fields will be more likely to be important  in the Synchrotron regime which is 
therefore the most relevant for the
observations. In this limit, the shape of the spectrum radiated by a relativistic particle 
is given by \citet{1986rpa..book.....R} 

\begin{equation}
\frac{dP}{d\nu}(\alpha,p,\nu)\,=\,\frac{\sqrt{3}q^3B}{m_ec^2}\sin{\alpha} F\left(\frac{\nu}{\nu_{\rm c}}\right),
\end{equation}

\noindent
where

\begin{equation}
F(x)\,=\,x\int_{x}^{±\infty}K_{5/3}(z)dz,
\end{equation}

\noindent
and  $K_{n}(z)$ is the modified Bessel function of second kind.   
Then the pitch angle averaged spectrum can be approximated as 

\begin{equation}
\frac{dP}{d\nu}(\gamma,\nu)\,=\,\frac{12\sqrt{3}\sigma_{\rm T}c U_{\rm B}}{\nu_{\rm L}}\sin{\alpha} G\left(\frac{\nu}{2\nu_{\rm c}^{*}}\right),
\end{equation}

\noindent
where

\begin{equation}
G(x)\,=\,x^2\left[K_{4/3}(x)K_{1/3}(x)-\frac{3x}{5}\left(K_{4/3}^2(x)-K_{1/3}^2(x)\right) \right].
\end{equation}

\noindent
In both cases, the spectrum increases with frequency 
$\frac{dP}{d\nu}\propto \nu^{1/3}$ up to the critical frequency 
$\nu_{c}$, (or $\nu_{c}^{*}\,=\,\frac{3}{2} \nu_{\rm L}\gamma^2$ in the 
angle averaged case), where there is a turn-over and an exponential 
cut-off. The emission is therefore peaked around the critical frequency, 
and both $\nu_{c}$ and $\nu_{c}^{*}$ scale like $B\gamma^2$. The 
condition of slow particle energy losses, given by equation\,\ref{eq:cond}, 
then implies a maximum frequency at which synchrotron emission can be 
produced.  The particles cannot radiate synchrotron emission above this 
energy because they cool before closing a full orbit. This maximum 
frequency depends only on fundamental constants and is independent of the 
magnetic field or particle energy: 
$h\nu_{\rm max}\,=\,3\,mc^2/\alpha_{f}\,=\,70$\,MeV. Any emission above 70\,MeV cannot be due to synchrotron.

\noindent
Of course in real situations the radiating particles are not 
mono-energetic but have a distribution in energy. One common case is a 
relativistic Maxwelian distribution: $N(\gamma)\propto \gamma 
\sqrt{\gamma^2-1}e^{-\gamma/\theta}$, where $\theta$ is the reduced 
temperature: $\theta\,=\,kT/m_e\,c^2$. In this case the spectrum peaks at 
$h\nu\,=\,1.5\,h\nu_{\rm L}\theta^2$, at lower frequencies the emitted 
spectrum is $\frac{dP}{d\nu}\propto \nu^{1/3}$ and an exponential a 
cut-off at higher energies. Another interesting case corresponds to a 
power-law particle energy distribution $N(\gamma)\propto \gamma^{-s}$, in 
the range $\gamma_{\min}<\gamma<\gamma_{\max}$.  Then the emitted 
spectrum is $\frac{dP}{d\nu}\propto \nu^{1/3}$ up to $\nu\,=\,1.5\,\nu_{\rm 
L}\gamma_{\rm min}^2$ where the photon spectral slopes changes to 
$\alpha\,=\,(s-1)/2$, ($\frac{dP}{d\nu}\propto \nu^{-\alpha}$) until it 
reaches a high frequency cut-off at $\nu\,=\,1.5\,\nu_{\rm L}\gamma_{\rm 
max}^2$.

\subsubsection{Synchrotron self-absorption}\label{sec:synchabs}

From the point of view of the electron, the emission of cyclo-synchrotron 
emission constitutes a spontaneous transition between energy levels. As 
in the case of atomic transitions, the spontaneous emission mechanisms 
coexist with the opposite transition which happens through absorption of 
radiation. The transition probabilities are described by the Einstein 
coefficients and their relations.  On can show (see e.g. 
\citet{1991MNRAS.252..313G} that the absorption coefficient (in 
cm$^{-1}$) can be expressed as a function of the emissivity 
$j_{\nu}\,=\,\int_{1}^{\infty}\frac{N(\gamma)}{4\pi} \frac{dP}{d\nu}d\gamma$ 
(erg s$^{-1}$\,Hz$^{-1}$\,cm$^{-3}$\,ster$^{-1}$):

\begin{equation}
\alpha_{\nu}\,=\,\frac{1}{2m_e\nu^2}\frac{1}{p\gamma}\frac{d \gamma p j_{\nu}}{d\gamma}.
\end{equation}

\noindent
where $\alpha_{\nu}$ gives the probability for photon travelling in the 
magnetised plasma to be absorbed per unit of distance crossed. The 
specific intensity $I_{\nu}$ (erg s$^{-1}$\,Hz$^{-1}$\,cm$^{-2}$\,ster$^{-1}$) 
of the synchrotron radiation escaping a homogeneous medium 
of size $L$ is given by the resolution of the radiative transfert 
equation

\begin{equation}
I_{\nu}\,=\,\frac{j_{\nu}}{\alpha_{\nu}}\left(1-e^{-\tau_{\rm sa}}\right),
\end{equation}

\noindent
where $\tau_{\rm sa}\,=\,\alpha_{\nu}L$ is the Synchrotron self-absorption optical
depth. There are two main assymptotic regimes depending on the value of
$\tau_{\rm sa}$. 
For $\tau_{\rm sa}<<1$ the medium is optically thin: $I_{\nu}\,\simeq\,j_{\nu}\,L $
while for $\tau_{\rm sa}>>1$  the medium is optically thick and he synchrotron
emission is self-absorbed: $I_{\nu}\,=\,j_{\nu}/\alpha_{\nu}$. As $\tau_{\rm sa}$
decreases with photon frequency, the low frequency part of the spectrum is
usually self-absorbed, while at higher frequency the emission is optically thin.
The transition between optically thin and thick emission occurs at a frequency
$\nu_{\rm t}$ such that by $\tau_{\rm sa}(\nu_{\rm t})\,\simeq\,1$. The value of 
$\nu_{\rm t}$  increases with  electron density (or Thomson depth $\tau_{\rm
T}\,=\,n_{\rm e}\,\sigma_{\rm T}\,L$). The exact value of the turnover frequency as
well as the  shape of the spectrum in the self-absorbed regime depends on the
shape of the electron energy distribution. 
For a thermal distribution the low frequency self-absorbed part of the spectrum
is a black body in the Rayleigh-Jeans regime i.e. $I_{\nu}\propto \nu^{2}$. For
a powerlaw energy distribution the turnover frequency can be estimated
analytically $\nu_{\rm t}\,\propto \,B^\frac{s+2}{s+4}\tau_{T}^\frac{2}{s+4}$ and
the low frequency self-absorbed spectrum is strongly inverted: 
$I_{\nu}\,\propto\,\nu^{5/2}$.

\subsection{Curvature radiation} 

Curvature radiation is emitted by relativistic particles moving along 
strongly curved magnetic field lines.  This process is important in the 
magnetosphere of pulsars where the field lines are curved and the 
magnetic field is so strong that the transverse component of the velocity 
is radiated very quickly and then the particles just follow very closely 
the magnetic field lines. In this case the acceleration is perpendicular 
to the fields lines and the emission is beamed in the direction of the 
trajectory.  The observer sees a pulse every time the line of sight 
intercepts the beam of radiation traveling with a nearly circular motion 
along the magnetic field lines.  This situation is in fact equivalent to 
synchrotron emission.  The radiated power and spectrum for one particle 
is therefore given by the same formulae as for synchrotron with the 
Larmor frequency replaced by the rotation frequency of the relativistic 
particles $\nu_{\rm L}\,=\,c/2\pi\,R$, where $R$ is the curvature radius of 
the field lines. The mono-energetic spectrum of curvature radiation has 
therefore the same form as that of synchrotron radiation. It varies like 
the cube root of frequency at low frequencies and falls exponentially 
above the peak frequency $\nu\,\simeq\,1.5\,\gamma^2\,c/2\pi\,R$ . Then the 
full observed spectrum depends on the energy and spatial distribution of 
the particles and magnetic fields in the pulsar magnetosphere. However, 
for likely values of the parameters, it turns out that the intensity 
produced by the incoherent sum of a single-particle curvature radiation 
is not enough to explain the very high brightness temperatures of the 
radio emission of pulsars.  It is necessary to assume that there are 
charged bunches containing $N\,\sim\,10^{15}$ electrons traveling together at 
the same velocity and concentrated in a same volume of dimension smaller 
than the emitted wavelength. Then, these electrons behave as a unique 
particle of charge $qN$. Since according to the Larmor formula the 
radiated power scales like the square of the charge of the particle 
(equation\,\ref{eq:larmorform}), the emission is amplified by a factor $N$ with 
respect to the sum of the individual emission of the $N$ particle. This 
amplification is due to the fact that the $N$ electrons radiate in 
coherence. Such coherence effects may be responsible for the observed 
emission from the magnetosphere of pulsars.

\subsection{Bremsstrahlung}

Bremsstrahlung is a radiation associated to the acceleration (or in this case
braking) of a free electron interacting with the electric field of an ion. The
ion is at rest and the electron, which travels with a velocity $v$, is
deflected. This deflection is associated with a transverse acceleration and a
distant observer sees a pulse of electric field. The duration of the pulse is of
the order of the duration of the interaction $\Delta t_{\rm int}\,\simeq\,b/v$,
where $b$ is the impact factor (i.e. the shortest distance at which the electron
approaches the ion). The photon spectrum is given by the power spectrum of the
pulse which, in the case of random incoherent pulses, is flat up to the cut-off
frequency $\nu_{\rm cut}\,\simeq\,(2\Delta t_{\rm int})^{-1}$ and exponentially
cut-off above that. The amplitude of the flat part of the spectrum for one
particle is

 \begin{equation}
 I_{\nu}\,=\,\frac{8Z^2e^6}{3\pi c^3m_{e}^2v^2b^2}.
 \end{equation}

\noindent
In a realistic situation we have to integrate this result not only on the 
energy distribution of the electrons but also on the distribution of 
their impact factors, since intensity and maximum energy of the spectrum 
for one particle both diverge when $b$ vanishes, an artificial minimum 
impact factor has to be assumed. This minimum impact factor can be 
estimated based on physical arguments, but the full answer requires a 
quantum physic treatment which introduces correction factors known as 
gaunt factors. the most common situation is the case of maxwellian 
electron energy distribution of temperature $T$. Assuming a fully ionised 
hydrogen plasma, the Bremsstrahlung emissivity is $j_{\nu}\propto n_{\rm 
e}n_{\rm p} T^{-1/2}\exp\left({-h\nu/kT}\right)$. The emitted spectrum is 
therefore flat up to a frequency corresponding to a photon energy 
comparable to the thermal energy of the electrons. The total power 
represents a cooling term for the plasma

 \begin{equation}
 J(T)\simeq 2.4 \times 10^{-27}\bar{g}_{ff}(T)n_{\rm e} n_{\rm p}  \sqrt{T}\quad{\rm erg\,s^{-1}\,cm^{-3}},
 \end{equation}

\noindent
where $\bar{g}_{ff}(T)$ is a gaunt factor of the order of unity. Since the
radiated power is proportional to square of the plasma number density,
Bremsstrahlung will be a very efficient cooling mechanism for dense plasma, and
much less efficient for diluted plasmas.  In a way that is very similar to
cyclo-synchrotron radiation, Bremsstrahlung emission can be self-absorbed. For a
thermal plasma, the emission is  Rayleigh-Jeans ($I_{\nu}\propto \nu^{2}$) 
below the turn over frequency.

 \subsection{Compton}
 \label{sec:compton}

Compton scattering plays an important role in the formation of the high 
energy emission of compact objects. In most elementary physics textbooks, 
it is described as the scattering of a photon by an electron at rest. 
This results for the photon in a change of direction of propagation and 
loss of energy to the electron. In the context of high energy 
astrophysics the electron can be energetic and moving at relativistic 
speeds. In this case the electron can transfer a significant amount of 
energy to the photon which may be up-scattered into the X-ray or even 
$\gamma$-ray domain. This process is then called inverse Compton. It is 
however exactly the same process as the usual Compton diffusion but 
considered in a different reference frame. The conservation of momentum 
and energy during the collision gives the general formula for the change 
of energy for the photon

 \begin{equation}
\frac{\nu'}{\nu}\,=\,\frac{1-\mu\beta}{1-\mu'\beta+\left(h\nu/\gamma m_ec^2\right)\left(1-\cos{\alpha}\right)},
\end{equation}

\noindent
where $\beta\,=\,v/c$ is the velocity of the electron as a fraction of the 
speed of light, $\mu$ and $\mu'$ are the cosine of the angle between the 
directions of the electron and photon momenta respectively before and 
after the interaction, $\nu$ and $\nu'$ are the photon frequencies 
respectively before and after the collision, $\alpha$ is the angle 
between the photon directions before and after interaction. If the photon 
is scattered by an electron at rest this formula simplifies to the 
classic formula for Compton exchange

\begin{equation}
\frac{\nu'}{\nu}\,=\,\frac{1}{1+\left(h\nu/m_e c^2\right)\left(1-\cos{\alpha}\right)}.
\end{equation}

\noindent
The photon looses energy in the process (electron recoil effect). If the initial
photon energy is small ($h\nu<<m_e c^2$)

\begin{equation}
\frac{\Delta\nu}{\nu}\,=\,-\frac{h\nu}{me c^2}(1-\cos\alpha).
\end{equation}

\noindent
If on the contrary, the electron is moving fast, the change in photon energy can
be seen as a Doppler effect caused by the change of reference frame. In the
electron rest frame the frequency of the incident photon is

\begin{equation}
\nu_{o}\,=\,\gamma\nu(1-\mu\beta),
\end{equation}

\noindent
and if $h\nu_{o}<<m_e c^2$, the change of photon energy in this frame is
negligible $\nu'_{o}\,=\,\nu_{o}$. Then going back to the lab frame, one gets

\begin{equation}
\nu'\,=\,\frac{\nu'_{o}}{\gamma(1-\mu'\beta)}\,=\,\frac{\nu_{o}}{\gamma(1-\mu'\beta)}
\,=\,\nu \frac{1-\mu\beta}{1-\mu'\beta}.
\label{doppler}
\end{equation}

\noindent
The maximum energy gain is obtained for a frontal collision with back scattering
of the photon.

\subsubsection{Cross section}

The number of scattering events per unit time and unit volume in the lab 
frame, for a mono-energetic beam of electrons interacting with a 
mono-energetic beam of photon is

\begin{equation}
\frac{dn}{dt}\,=\,n_e n_\nu \sigma_{KN} c \left(1-\beta\mu\right), 
\label{eq:rate}
\end{equation}

\noindent
where $n_{e}$ and $n_{\nu}$ are the respective electron and photon 
densities and $\beta$ the reduced speed of the electrons and $\mu$ the 
cosine of the angle between the two beams, as above. $\sigma_{KN}$ is the 
Klein-Nishina the cross section

\begin{equation}
\sigma_{KN}\,=\,\frac{2\pi r^{2}_{e}}{x}\left[\left(1-\frac{4}{x}-\frac{8}{x^{2}}\right)
\ln\left(1+x\right)+\frac{8}{x}-\frac{1}{2(1+x)^{2}}+\frac{1}{2}\right],
\end{equation}

\noindent
where $x$ 

\begin{equation}
x\,=\,\frac{2h\nu}{m_ec^2}\gamma(1-\mu\beta).
\end{equation}

\noindent
and $r_{e}\,=\,e^{2}/m_e\,c^2$ is the classical electron radius. In the 
non-relativistic limit ($x<<1$), the Klein-Nishina cross-section reduces 
to the Thomson cross-section: $\sigma_{KN}\,\simeq\,8\pi\,r^{2}_{e}/3\,=\,\sigma_{T}$. This non-relativistic limit corresponds to the 
Thomson regime. Around x$\,\sim\,1$ the cross section starts to decrease 
with increasing $x$, in the ultra-relativistic regime: $x>>1$, the cross 
section decreases as

\begin{equation}
\sigma_{KN} \simeq\frac{3}{8} \frac{\sigma_T}{x} \left[1+2\ln(1+x)\right].
\end{equation}

\noindent
This strong reduction makes inverse Compton highly inefficient in the 
Klein-Nishina regime. Finally, the Compton cross section scales like 
$m_{\mathrm{e}}^{-2}$. The scattering cross section on charged particles 
other than an electron or positron are therefore smaller by a factor of 
at least $(m_{\mathrm{e}}/m_{\mathrm{p}})^{2}\,\sim\,10^{-7}$. These 
processes are therefore completely negligible.

\subsubsection{Radiated power and amplification}

In the Thomson regime ($\gamma h\nu<< m_e c^2$) and neglecting the recoil 
effect ($p^2>>h\nu/m_ec^2$) the total power radiated by one electron in 
the field of soft photons of energy density $U_{\rm ph}$ is 
\citep{1986rpa..book.....R}:

 \begin{equation}
 P\,=\,\frac{4}{3}\sigma_{\rm T}c p^2 U_{\rm ph},
 \label{eq:comptonpow}
 \end{equation}

\noindent
where $p\,=\,\gamma\beta$ is the reduced momentum of the electron and $U_{\rm 
ph}$ is the energy density of the target photon field. It is interesting 
to note that the radiated power in this limit is independent of the 
actual energy of the soft photons (which do not have to be 
mono-energetic), or their spectrum.

\noindent
From equation\,\ref{eq:comptonpow} and \ref{eq:rate}, the average photon 
energy amplification in the course of one interaction is given
by

\begin{equation}
\frac{\langle\nu'\rangle}{\nu}\,=\,1+\frac{4}{3}p^2.
\label{eq:compa}
\end{equation}

\subsubsection{Spectrum from single scattering of a power-law energy distribution of relativistic 
electrons}

Let us consider an isotropic distribution of photons of typical energy 
$\nu_0$, scattering off a power-law energy distribution of electrons 
(i.e. $N(\gamma)\,=\,N_0\gamma^{-s}$ in the range $\gamma_{\rm 
min}<\gamma<\gamma_{\rm max}$), in the Thomson regime.  After interaction 
with an electron of momentum $p$, the photons have an average energy: 
$\nu\,=\,\nu_0(1+4 p^2/3)\,\simeq\,\nu_0\,4\,\gamma^2/3$.  As a first approximation 
we can neglect the dispersion of the scattered photon energies and assume 
that all the power radiated by the electrons of energy $\gamma$ is 
emitted at the average scattered photon energy $\nu$. The emitted 
spectrum (erg\,s$^{-1}$\,$Hz^{-1}$\,cm$^{-3}$) is then

\begin{equation}
j_{\nu}(\nu)\,=\,\frac{4}{3}\sigma_{\rm T}c p^2 U_{\rm ph}N(\gamma) \frac{d\gamma}{d\nu}\simeq \frac{2}{3}\sigma_{\rm T}cU_{\rm ph}\frac{N_0}{\nu_0}\left(\frac{3\nu}{4\nu_0}\right)^{-\left(\frac{s-1}{2}\right)},
\end{equation}

\noindent
for $\nu$ comprised between $\nu_{\rm min}\,\simeq\,4\nu_0\gamma_{\rm min}^2/3$ and
$\nu_{\rm max}\,\simeq\,4\nu_0\gamma_{\rm max}^2/3$. The result is therefore a
power-law ($j_{\nu}\propto \nu^{-\alpha}$) extending up to $\nu_{max}$ with a
slope $\alpha\,=\,(s-1)/2$ which is identical to that obtained in the case of
synchrotron radiation.  If instead,  the electrons at  $\gamma_{max}$ emit in
the Klein-Nishina regime, the maximum energy of the Compton radiation is limited
by the energy of the electrons $h \nu_{\max}\,=\,\gamma_{\rm max}m_{e}c^2$. However
due to the sharp decrease of the cross section in the KN regime the spectrum
deviates from a power-law and is strongly suppressed at photon frequencies that
are above  $\nu_{\rm KN}\,\simeq\,m_e^2c^4/(h^2\nu_0)$.

\subsubsection{Multiple Compton Scattering: Thermal Comptonisation}
\label{sec:thcomp}
  
In a realistic situations the photons can undergo zero, one, or several
successive scattering before escaping from the energetic electron cloud. In the
Thomson regime the probability of interaction is determined by the Thomson
optical depth: $\tau_{\rm T}\,=\,n_{e} \sigma_T R$, where $n_{e}$ is the electron
number density and $R$ is the typical dimension of the medium. The probability
for a photon to cross the medium without scattering any electron is
$\exp(-\tau_{\rm T})$. In the optically thin case ($\tau_{\rm T}<1$), and
$\tau_{\rm T}$ represent the average number of scatterings before escape (or, in
other words its probability of interaction). For $\tau_{\rm T}>1$ the photons
interact at least once and the average number of scattering before escape is
$\sim\,\tau_{\rm T}^2$. Multiple Compton scattering implies non-negligible
optical depth, and since the emitting region in compact objects is necessarily
small, this also imply a significant density.  If the density is large, most of
the particles are thermalised. So the most relevant case in which multiple
Compton scattering is important is the case of soft photons up-scattered in a
hot thermal plasma of electrons (i.e. the electrons have a Maxwellian energy
distribution).  This process is called thermal Comptonisation. The Comptonized
emission depends not only on the Thomson depth but also on the temperature
$T_{e}$  of the electrons.  As long as the photon energy $h\nu$ is smaller than
the typical energy of the electrons $\sim\,kT_{e}$, the photons gain energy at
each interaction. In the Thomson limit the average fractional energy gain can be
approximated as

\begin{equation}
\frac{\Delta\nu}{\nu}\,=\,4 \theta_{\rm e}+16 \theta_{\rm e}^2,
\end{equation}

\noindent
where $\theta\,=\,kT_{\rm e}/m_{\rm e} c^2$  \citep{1986rpa..book.....R}.

\noindent
Usually one defines the Compton $y$ parameter as the product of this average fractional energy 
exchange times the average number of interaction:

\begin{equation}
y\simeq\left(4 \theta+16\theta^{2}\right)\left(\tau_{\rm T}+\tau_{\rm T}^2\right).
\end{equation}

\noindent
Then one can show that the energy of the incident photon is amplified by a
factor

\begin{equation}
A\,\simeq\,e^y\left[{1+\left(e^y-1\right)\frac{h\nu_0 }{m_e c^2}\left(4 \theta+16\theta^2\right)^{-1}}\right]^{-1}.
\end{equation}

\noindent
In the limit of small $y$, the amplification factor reduces to $A\,\simeq\,1+y$.
While in the limit of large $y$, the Comptonisation process is said to be
saturated and most of the radiation is emitted at an energy $\sim\,(4 \theta+16
\theta^2)m_{\rm e} c^2$, that is comparable to average energy of the electrons.

\noindent
The escaping spectrum is constituted of the sum of contributions from the
photons that have escaped without interaction, plus those that have interacted
once before escaping, plus those that have interacted twice, trice...etc. Each
of these spectral components,  which are called  `Compton orders', forms a broad
hump. The average photon energy of each  hump increases with the number of
interactions until it becomes  comparable to the thermal energy of the
electrons. These humps are however apparent in the spectrum only in the
optically thin regime $\tau_{\rm T}<<1$.  In practice, the observed $\tau_{\rm
T}$ are often of the order of unity or larger. In this case the Compton orders
are too close and too blended to be individually distinguishable. They combine
to form power-law spectrum  extending from the typical energy of the seed
photons up to the energy of the thermal electrons \citep{1980A&A....86..121S}.
At higher energies the spectrum drops-off exponentially. The photon index
$\Gamma$ (defined as $F_{\nu}\propto \nu^{1-\Gamma}$) can be approximated as

\begin{equation}
\Gamma \simeq C (A-1)^{\delta},
\end{equation}

\noindent
with $C\,\simeq\,2.33$ and $\delta\,\simeq\,1/6$ depending weakly on the temperature
of the electrons, the average energy of the seed photons and the exact geometry
of the comptonizing plasma \citep{1999ASPC..161..295B}.  For instance, for electron
temperatures in the range 50--200\,keV as mostly observed in X-ray binaries, and
for a black body seed photons of temperature $kT_{bb}$\,=\,0.15\,keV, numerical
simulations give $C\,=\,2.19$, $\delta\,=\,2/15$ \citep{2001MNRAS.326..417M}.

\noindent
By modelling observed Comptonized spectra it is possible to infer the
temperature of the electrons and the Thomson depth. Note that the X-ray spectral
slope $\Gamma$ depends on both parameters (mostly via $y$). There is therefore a
degeneracy which can be broken only if one observe simultaneously at higher
energies and measure the Comptonisation cut-off which is usually found around
100\,keV (see Section\,\ref{sec:bhbaccretion}).

\begin{figure} 
\includegraphics[scale=0.35]{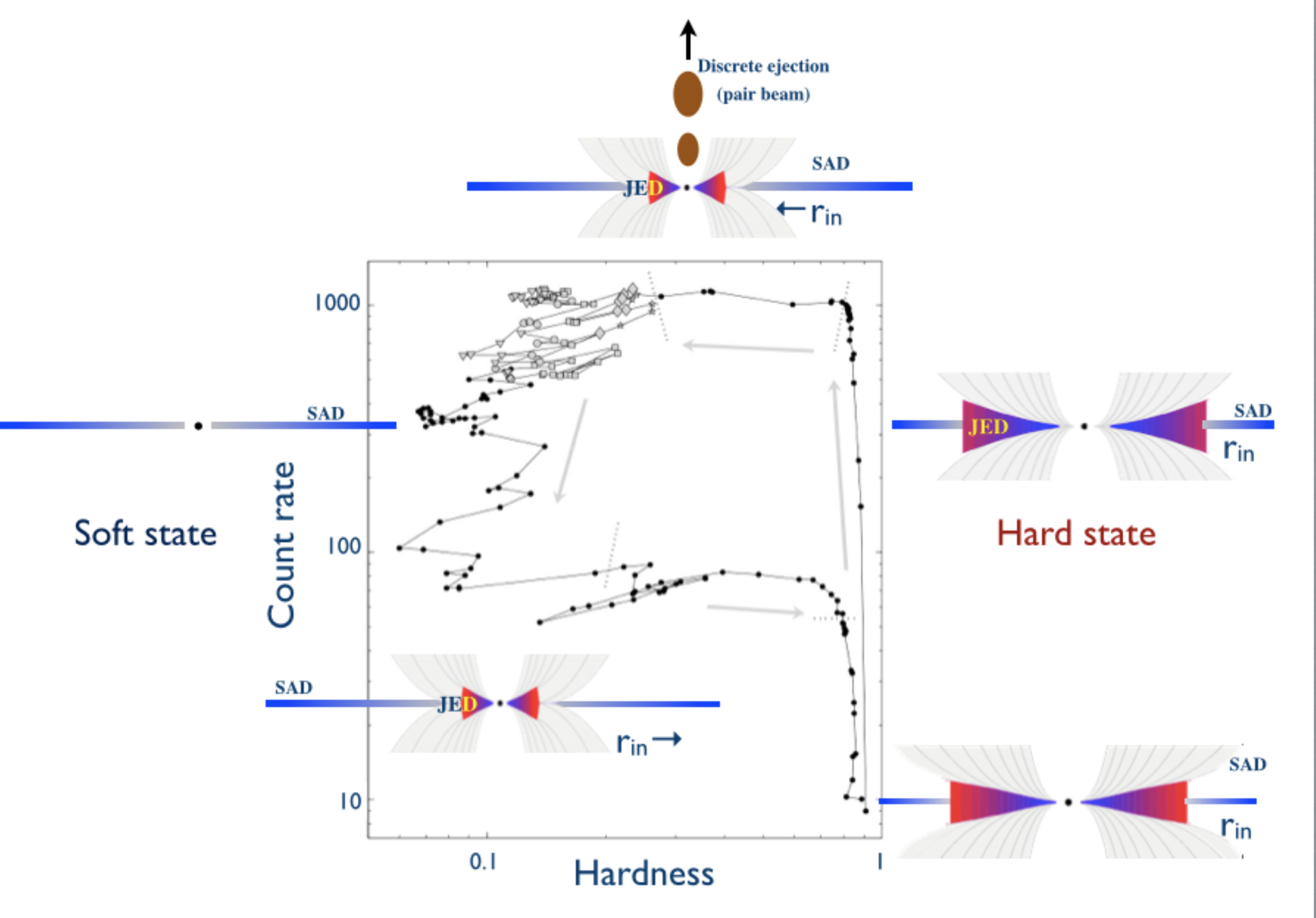} 
\caption{Typical track followed by the BHB GX\,339--4 in the HID diagram during an
outburst.  The various sketches illustrate a possible scenario for the evolution
of the geometry of the accretion flow in the context of a truncated disc model
in which the central hot accretion flow takes the form of a jet emitting disc 
(Courtesy: P.O Petrucci). }
\label{fig:fig1} 
\end{figure}

\begin{figure*}
\center{\includegraphics[scale=0.3]{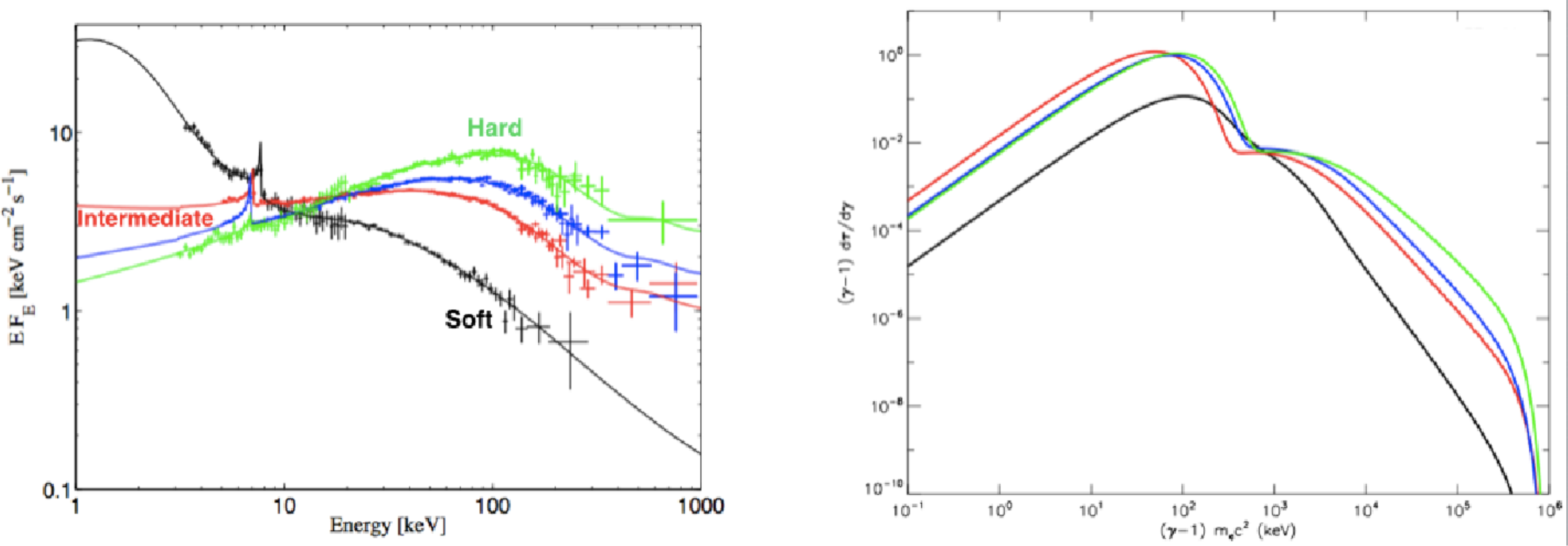}}
\caption{Left: Joint INTEGRAL/JEM-X, IBIS and SPI energy spectra of Cyg\,X--1
during four different spectral states fitted with the magnetised hybrid
thermal-non-thermal Comptonisation model BELM. Right: Energy distribution 
of the Comptonizing  electrons obtained in the best fit models of the 
left-hand side panel. These fits set an upper limit on the 
amplitude of the magnetic field in the X-ray corona at about $10^{5}$\,G  
in the harder states and $10^{7}$\,G 
in softer states \citep{2013MNRAS.430..209D}.}
\label{fig:fig2}
\end{figure*}

\section{Model for the accretion flow}
\label{sec:bhbaccretion}

Accreting BHs in X-ray binary systems (hereafter BHBs) produce strongly variable
radiation over the whole electromagnetic spectrum. From the radio to IR band,
non-thermal emission is often detected and usually associated to synchrotron
emission from very high energy particles accelerated in jets   (see e.g.
\citealt{2003MNRAS.346..689C}, hereafter C03; \citealt{2011ApJ...740L..13G},
hereafter G11). The same particles can occasionally produce detectable emission
in the GeV range \citep{2009Sci...326.1512F, 2009Natur.462..620T,
2013MNRAS.434.2380M}. The thermal emission from the accretion disc can be
studied in the optical to soft X-ray bands. In the hard X-ray domain (above a
few keV) the emission is dominated by a non-thermal component  which nature and
origin is strongly debated, and which is associated to the emission of a hot and
tenuous  plasma located in the direct environment of the BH. This plasma is
generically called the `corona', drawing analogy from the solar corona, although
the formation of the BH corona,  its power feeding and emission processes are
most likely very different from that of the Sun. BHBs also show strong aperiodic
and quasi-periodic time variability down to the millisecond scale in X-rays,
optical, and IR bands  (see e.g. \citealt{2010MNRAS.407.2166G,
2008MNRAS.390L..29G}  herafter C10 and G08 respectively).  They constitute prime
targets for HTRA. 

Most BHBs are transient X-ray sources detected during outbursts lasting 
from a few months to a few years. During outbursts their luminosity can 
increase by many orders of magnitude to reach values close to the 
Eddington limit ($L_{\rm E}\,\simeq\,10^{39}$\,erg/s for a 10\,$M_{\odot}$ 
BH), before going down, back to quiescence. These sources therefore 
constitute a unique laboratory to investigate how the physics of the 
accretion and ejection depends on the mass accretion rate onto the BH. 
During these outbursts, not only the luminosity, but also the broad band 
spectral shape changes drastically as a source evolves through a 
succession of X-ray spectral states, showing very different spectral and 
timing properties. Nevertheless those X-ray spectral states can all be 
described in terms of the same spectral components arising from the 
thermal accretion disc, X-ray corona and jets.  Their rich timing 
phenomenology (see Chapter\,4 by T.\,Belloni in these proceedings) is poorly 
understood. In this section I will focus on a model that can explain 
qualitatively most of the timing and spectral feature arising from the 
accretion flow.


\subsection{Spectral states}
 
The spectral evolution during outburst is usually studied using Hardness 
Intensity Diagrams (HID). One of such diagrams is shown in 
Figure\,\ref{fig:fig1}. There are two main stable spectral states, namely 
the soft and the hard state, corresponding respectively to the left and 
right hand side vertical branches of the HID (see Figure\,\ref{fig:fig1}). 
The other spectral states are mostly short-lived intermediate states 
associated to transitions between the two main spectral states. Some 
X-ray spectra of Cyg\,X--1 in its soft (in black), hard (green) and 
intermediate (blue and red) states are shown in Figure\,\ref{fig:fig2}.

The soft spectral state is observed at luminosity levels ranging 
approximately from $10^{-2}$ to a few $0.1 L_{\rm E}$. In this state the 
high energy emission is dominated by the soft thermal multi-blackbody 
disc emission peaking around 1\,keV, for this reason this state is also 
called `thermal dominant' by some authors \citep{2006ARA&A..44...49R}. 
The intense thermal radiation and hot disc temperature ($\sim\,$\,1\,keV) is 
consistent with that of a standard geometrically thin, optically thick 
accretion disc \citep{1973A&A....24..337S} extending down very close to the 
BH. The coronal emission is usually very weak forming a non-thermal 
power-law tail above a few keV. The geometry of the corona is 
unconstrained in this state but it is generally assumed to be constituted 
of small-scale magnetically active regions located above and below the 
accretion disc \citep{1979ApJ...229..318G}. In these regions the 
energetic electrons of the plasma up-scatter the soft X-ray photons 
coming from the disc into the hard X-ray domain.  Due to the weakness of 
the non-thermal features, the soft state is perfect to test accretion 
disc models and measure the parameters of the inner accretion disc. In 
particular the detailed model fitting of the thermal emission of the 
accretion disc indicates that the inner radius of the disc is a constant 
and is independent of the luminosity of the system 
\citep{2004MNRAS.347..885G}. This constant inner radius is believed to be 
located at the innermost stable circular orbit (ISCO) that is predicted 
by the theory of general relativity and below which the accreting 
material must fall very quickly across the event horizon. As the size of 
the ISCO is very sensitive to the spin of the BH this offers an 
opportunity to constrain the spin of BHs in X-ray binaries. This may also 
be used to validate the spin measurements made using the relativistic 
iron $K_{\alpha}$ line profile which remain the only method that can be 
used in AGN. This is a difficult task because measuring the disc inner 
radius accurately requires very detailed disc emission models taking into 
account the general relativistic effects as well as non-blackbody effects 
through detailed disc atmosphere models 
\citep{2005ApJ...621..372D,2006ApJ...647..525D}. This method also 
requires the knowledge of the distance and inclination of the system.  
Nevertheless the most recent BHBs spin estimates obtained from disc 
continuum and line fitting are converging (see 
\citet{2016ASSL..440...99M} for a recent review of these issues).  
Although the soft state is associated to strong Fe absorption lines 
indicative of a disc wind \citep{2014MNRAS.444.1829P}, there is no 
evidence so far of any relativistic jet component in soft state 
\citep[see][]{2011ApJ...739L..19R}.

The hard state is observed at all luminosities up to a few 0.1 $L_{E}$. 
In this state the emission from the accretion disc is much weaker and 
barely detected, the inferred temperature of the inner disc is also lower 
($T_{\rm in}\sim\,$0.1\,keV) than in soft state state 
\citep{2009MNRAS.396.1415C, 2011MNRAS.411..337D, 2015A&A...573A.120P}. 
The X-ray emission is dominated by a hard power-law with photon index 
$\Gamma$ in the range 1.5--2.1, and a high energy cut-off around 50--200\,keV. This kind of spectra is very well represented by Thermal 
Comptonisation models with Thomson depth 1-3 and electron temperatures in 
the range 20--200\,keV. In addition there is evidence for a Compton 
reflection component: the illumination of the thin accretion disc by the 
Comptonized radiation lead to the formation of a broad hump peaking 
around 30\,keV in the high energy spectrum and a prominent line around 
6.4\,keV caused by iron fluorescence (F$_{\rm e}$ K$_{\alpha}$) line. The 
amplitude $R$ of the reflection component is defined relatively to the 
reflection produced by an isotropic X-ray source above an infinite slab. 
In the hard state, $R$ is usually small, $R< 0.3$.  Another 
characteristic of the hard ste is the presence of radio emission with a 
flat of weakly inverted spectrum that is the signature of steady compact 
jets. Such compact jets are probably the most common form of jets in 
X-ray binaries, they appear to be present in all BH and NS binaries when 
in the hard X-ray spectral state. Their SED
extends from the radio to the mid-IR (e.g. 
\citet{2000MNRAS.312..853F}, \citet{2002ApJ...573L..35C}; hereafter C03, 
\citep{2010ApJ...710..117M}.

The measurements of the hard state X-ray spectrum put interesting 
contraints on the geometry of the corona in the hard state.  For 
instance, an isotropic corona made of active regions located above the 
disc (as that envisioned for the soft state) would produce significantly 
stronger reflection features ($R\sim\,1$ ) which may also be enhanced by 
general relativistic light bending \citep{2004MNRAS.349.1435M}. In 
addition, most of the illuminating radiation would be absorbed in the 
disc (typical disc X-ray albedo is $\sim\,0.1$) and re-emitted at lower 
energy in the form of nearly thermal radiation. This geometry would thus 
imply a strong thermal component with an observed flux that would be at 
least comparable to that of the corona.  If the corona has a significant 
Thomson depth, as inferred by the observations ($\tau_{\rm T}\ge 1$) and 
is radially extended above the disc, the reflection and reprocessing 
features might be smeared out by Compton scattering 
\citep{2001MNRAS.328..501P}. However detailed Monte-Carlo simulations 
have shown that in this case the strong reprocessed emission from the 
disc illuminating the corona would then cool down its electrons (via 
inverse Compton) to temperatures that are much smaller than observed 
\citep{1993ApJ...413..507H, 1994ApJ...432L..95H, 1995ApJ...449L..13S, 
2001MNRAS.326..417M} and then much steeper X-ray spectra would be 
observed.  In fact, the high electron temperature and weakness of the 
reflection/and disc features suggest that the corona and the disc see 
each other with a small solid angle.

In this context, a geometry that is favoured is that of the truncated 
disc model, where the accretion disc does not extend down to the ISCO, 
but is truncated at some larger distance from the BH. As the disc does 
not extend deeply into the gravitational potential of the BH, its 
temperature is lower. As a consequence the blackbody emission is much 
weaker than in soft state. The corona is constituted of a hot 
geometrically thick accretion flow that fills the inner hole of the 
accretion disc \citep{1997ApJ...489..865E, 1997MNRAS.292L..21P}. This hot 
flow Comptonises the soft photons from the disc and/or internally 
generated synchrotron photons to produce the hard X-ray continuum. The 
outer disc receives little illumination from the corona and produces only 
weak reflection and reprocessing features. This scenario explains 
qualitatively many of the spectral and timing properties of BHBs in the 
hard state, such as the correlation between X-ray spectral slope and 
reflection amplitude \citep{1999MNRAS.303L..11Z} or Quasi-Periodic 
Oscillations (QPO) frequencies simply by assuming that the disc inner 
radius changes for instance with luminosity 
\citep[see]{2007A&ARv..15....1D} The location of the disc truncation 
radius could be determined by a disc evaporation/condensation equilibrium 
\citep{2000A&A...361..175M, 2012ApJ...744..145Q}. Figure\,\ref{fig:fig1} 
shows a sketch of the possible evolution of the geometry of the accretion 
flow during an outburst.

Observationally, however, due to the weak disc features the actual 
transition radius is very difficult to measure accurately. The most 
recent estimates suggest that in the bright hard state, the inner disc is 
actually at a few gravitational radii at most \citep{2015ApJ...799L...6M, 
2015ApJ...808....9P, 2015MNRAS.451.4375F}. These results question whether 
the disc is actually truncated at all. But there is evidence that the 
disc recedes at low luminosity and the truncation radius could be located 
much farther out \citep[see][]{2015A&A...573A.120P}.

At least at high luminosities, the hot flow of the truncated disc model 
cannot be a standard advection-dominated accretion flow (ADAF; 
\citealt{1997ApJ...478L..79N}).
Indeed, ADAF theory does not predict hot flows 
that are as bright and as optically thick as observed ($\tau_{\rm T} > 
1$) in the bright hard state.  The large Thomson depth implies a large 
density which causes too much cooling for a hot solution to exist at 
$\tau_{\rm T}>1$. A possible fix to the model was proposed by 
\citet{2010ApJ...712..639O, 2012PASJ...64...15O}. It consists in assuming 
the presence of strong magnetic fields in the hot flow.  The magnetic 
field is strong enough to support the hot flow with magnetic pressure 
rather than the usual thermal pressure. This increases the scale height 
of the flow and decrease its density, allowing for solutions with larger 
$\tau_{\rm T}$. In this context, the Thomson depth, electron temperature 
and coronal luminosity observed in a typical hard state of Cyg\,X--1 
require the magnetic pressure to be a few times larger than thermal 
pressure \citep{2009MNRAS.392..570M}.

\subsection{Non-thermal particles}

In the soft state there are indications (at least in the prototypical 
source Cyg\,X--1) that the hard X-ray emission extends as a power law at 
least up to a few MeV \citep{2002ApJ...572..984M}. The absence of cut-off 
below 1 MeV indicates that the Comptonizing electron distribution of the 
soft state corona cannot be purely thermal. Producing such gamma-ray 
emission through inverse Compton requires a power-law like distribution 
of electrons extending at least up to energies of order of 10-100 times 
the electron rest mass energy.  In the hard state, an excess with respect 
to a pure Comptonisation model is detected in all bright sources 
\citep[e.g.][]{2007ApJ...657..400J,2010ApJ...717.1022D,2012ApJ...761...27J} 
which is also interpreted as the signature of a population of non-thermal 
electrons in the corona.

These findings triggered the development of hybrid thermal/non-thermal 
Comptonisation models. In these models the Comptonizing electrons have a 
similar energy distribution in all spectral state i.e. a Maxwellian with 
the addition of a high energy power-law tail \citep{1998PhST...77...57P, 
1999ASPC..161..375C}. These models have been extremely successful at 
fitting the broad band high energy spectra of BHBs. Figure\,\ref{fig:fig2} 
shows examples of INTEGRAL spectra fit with an hybrid model. The 
transition from mostly thermal (in hard state) to mostly non-thermal (in 
soft state) emitting electrons is understood in terms of the radiation 
cooling. As a source evolves towards the soft state the corona intercepts 
a much larger flux of soft photons from the disc. In this more intense 
soft radiation field the Compton emission of the hot electrons of the 
plasma is stronger. They radiate their energy faster. This makes the 
electron temperature significantly lower in softer states. As a 
consequence, the Compton emissivity of the thermal particles is strongly 
reduced in the soft state and the emission becomes dominated by the 
higher energy non-thermal particles. In addition the Thomson depth of 
thermal electrons is found to be smaller in soft state, possibly because 
most of the material in the corona has condensed into the disc or is 
ejected during state transition. This further decreases the luminosity of 
the Maxwellian component of the plasma, making it barely detectable in 
soft state. Attempts to model the spectral evolution during spectral 
state transitions have confirmed that the huge change in the flux of soft 
cooling photon from the disc illuminating the corona drives the spectral 
changes of the corona \citep{2008MNRAS.390..227D}.

The most recent version of the model also includes the radiative effects 
of magnetic field on the lepton energy distribution 
\citep{2008A&A...491..617B,2009ApJ...698..293V}. Internally generated 
synchrotron photons (typically in the optical/UV range) constitute a 
source of seed photons for the Comptonisation process. In addition, the 
process of Synchrotron self-absorption provides an efficient coupling 
between leptons which can quickly exchange energy by rapid emission and 
absorption of synchrotron photons leading to very fast thermalisation of 
the lepton distribution on time-scales comparable to the light crossing 
time. In fact, it is not necessary to assume the presence of thermal 
Comptonizing electrons in the first place. The heating mechanism could be 
purely non-thermal e.g. accelerating all electrons into a power-law 
energy distribution. The thermalising effects of the so-called 
synchrotron boiler (in addition to Coulomb collisions) naturally leads to 
an hybrid thermal/non-thermal particle energy \citep{2009MNRAS.392..570M, 
2009ApJ...690L..97P}.

One important effect of the presence of non-thermal electrons in the 
corona in presence of magnetic field is the production of a strong 
synchrotron radiation component. The spectral break associated to the 
transition from optically thick to optically thin synchrotron emission is 
expected to be located in the optical/IR range. Thermal electrons also 
produce synchrotron radiation but at a much lower level. 
\citet{2001MNRAS.325..963W} have shown that the addition of a small 
fraction of non-thermal electrons in the corona (carrying a fraction $<$ 
10 \,\% of the total internal energy of the distribution) can increase the 
optical synchrotron flux by several orders of magnitude.  In the 
framework of the truncated accretion flow model, the hybrid thermal 
non-thermal distribution of electrons may dominate the optical and IR 
flux (in hard state) through synchrotron emission, while at the same time 
explain the hard X-ray through Comptonisation 
\citep{2013ApJ...778..165V}.

\subsection{X-ray timing}

X-ray binaries harbour a strong rapid X-ray variability which presents a 
very complex and richly documented phenomenology (see Chapter\,4 by  
T.\,Belloni these proceedings).  On the other hand, observations show that most 
of the X-ray variability occurs on time-scales ranging from 0.1 to 10 s. 
Strong variability is also observed on longer time-scales. In persistent 
sources, it was possible to construct long term X-ray fourier power 
density spectra (PDS) showing a nearly $1/f$ noise extending up to time 
scales of years \citep{2010LNP...794...17G}. In comparison there is 
virtually no variability on time-scales shorter than 0.01\,s.  The main 
problem that any model must overcome is a time-scale problem. Indeed, 
most of the high energy photons (and variability) must originate deep in 
the potential from a region of size $R<100 R_{\rm G}\,\sim\,$\,1500\,km.  The 
times scales in this X-ray emitting region are controlled by dynamical 
time-scales of the accretion flow such as the Keplerian time-scale

\begin{equation}
t_{\rm K}\,=\,0.3 \left(\frac{M}{10M_\odot}\right)\left(\frac{R}{50 R_{\rm G}}\right) {\qquad \rm s},
\end{equation}

\noindent
or the viscous time-scale (comparable to the accretion time-scale)

\begin{equation}
t_{\rm vis}\,=\,\left(\frac{H}{R}\right)^{-2} \frac{t_{\rm K}}{2\pi\alpha},
\end{equation}

\noindent
where $\alpha$ is the usual viscosity parameter $\alpha\,\sim\,0.1$ (see
Section\,\ref{sec:kepad}). In the case of a thin gas pressure dominated accretion
disc: $H/R\,\sim\,10^{-2}$ and $t_{\rm vis}\,\sim\,10^4 t_{\rm K}\,\sim\,10^3$ s, while
in the case of a hot flow: $H/R\,\sim\,0.3$ an $t_{\rm vis}\,\sim\,10\,t_{\rm K}\,\sim\,1-10$\,s.  
This means that high Fourier frequencies can easily be produced in
this region. But the longest observed times scales are too long to be produced
in the region of main energy release. They must be generated in the outer parts
of the accretion flow.

A possibility could be that fluctuations are generated at large distance 
and propagate inward. For instance, \citet{1997MNRAS.292..679L} postulates 
fluctuations in viscosity on local inflow time-scale (itself viscous) over 
a wide range of distances $R$ from the BH and computes the resulting 
fluctuations of the mass accretion rate at the inner disc radius. For 
fluctuation amplitudes that are independent of $r$, this gives a power 
spectrum $p( f ) \propto 1/ f$ i.e. slow variations of large amplitude.  
In this model the amplitude of the fluctuations generated at a given 
radius is modulated by the longer fluctuations propagating from the outer 
region of the disc.  For this reason a linear correlation is expected 
between the the average flux measured within a given time interval and 
the root-means-squared (RMS) amplitude of the variability during the same time-interval.  Such 
RMS-flux correlations are ubiquitous among accreting BH sources and 
difficult to produce by any other model (see discussion in 
\citet{2005MNRAS.359..345U}..  Another intriguing feature of the rapid 
variability of BH binaries is the existence of Fourier frequency 
dependent delays between X-ray energy bands (the hard photons lagging 
behind the softer photons).  \citet{2001MNRAS.327..799K} improved upon 
Lyubarskii's model and showed that assuming a harder spectrum close to 
the BH produces a logarithmic dependence of time-lags on energy, as 
observed.

   \begin{figure*}
   \includegraphics[scale=0.28]{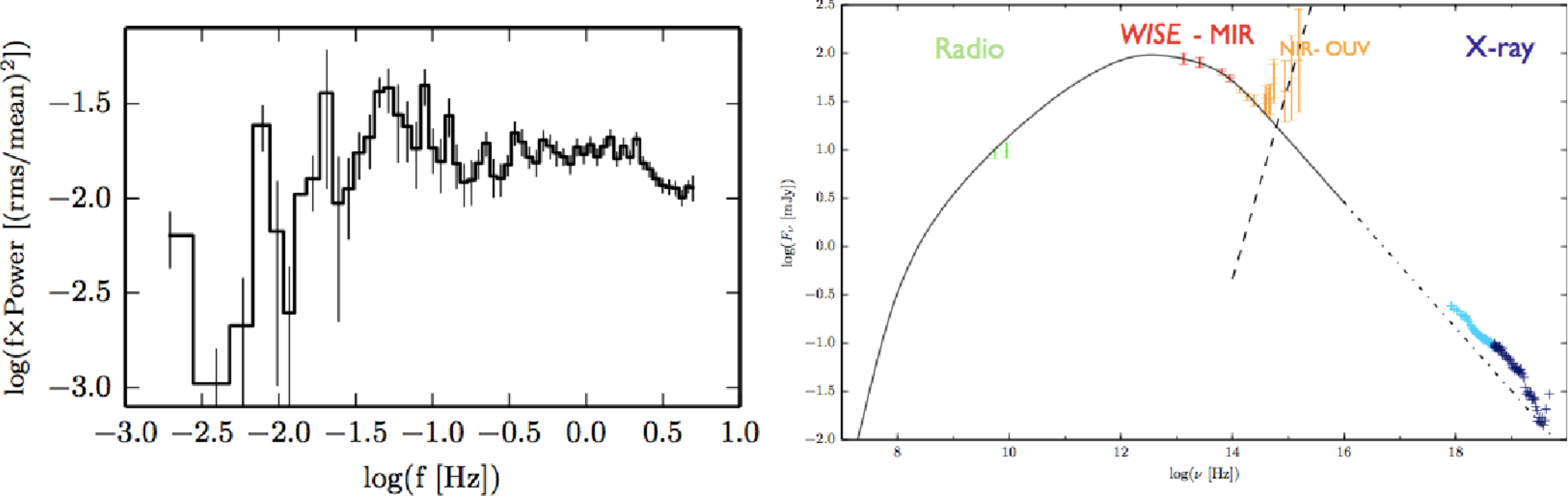}
 \caption{Left: the observed X-ray PSD of GX\,339--4 during the  observations
presented in G11. Right: the SED measured by G11 compared to a simulated jet SED
obtained assuming that the jet Lorentz factor fluctuations have exactly the PSD
as the X-ray flux \citep{2015MNRAS.447.3832D}.}
 \label{fig:1bis}
 \end{figure*} 

The shape and amplitude of the power spectrum changes drastically with spectral
state (again, see Chapter\,4 by T.\,Belloni in these proceedings). In the soft state the
RMS amplitude of X-ray variability  is at most a few percent, while in the hard
state  it can reach 30 \,\%. In the hard state the PDS shows band-limited noise.
An example of observed  X-ray PDS in the hard state is displayed in 
Figure\,\ref{fig:1bis}. It can be approximatively represented as flat, up to a
break  frequency $\nu_{b}\,\sim\,0.01$\,Hz  at which the slope changes. Above
$\nu_{\rm b}$,  the power decreases approximatively as $1/f$ up to a second 
break at a frequency $\nu_{h}\,\sim\,$ a few Hz. Above $\nu_{h}$, the PDS decreases
now like $f^{-2}$ (or steeper). The observed band limited noise PDS are usually
well  described in terms of the sum of 4-5 broad Lorentzians  (e.g. Nowak 2000;
van der Klis 2006). In the truncated disc scenario the band limited noise
variability constitutes the variability that is generated due to propagating
fluctuations in the hot flow.  $\nu_b$ is naturally associated as the viscous
frequency at the disc inner radius and $\nu_h$ is the viscous time-scale at the
inner radius of the hot flow which may differ from the ISCO 
\citep{2007A&ARv..15....1D,2011MNRAS.415.2323I,2015ApJ...807...53I}.

In addition, a low frequency QPO (LFQPO) is often observed at some 
intermediate frequency between $\nu_{b}$ and $\nu_{l}$.  This LFQPO is 
widely believed to be caused by Lense-Thirring (LT) precession of the hot 
flow (Ingram, Done \& Fragile 2009). LT precession is a frame dragging 
effect associated to the misalignment of the angular momentum of an 
orbiting particle and the BH spin, leading to precession of the orbit. 
Numerical simulations have shown that in the case of a hot geometrically 
thick accretion flow, this effect can lead to global precession of the 
hot flow \citep{2007ApJ...668..417F}. The hot flow precesses like a solid 
body, and the precession frequency is given by a weighted average of the 
LT precession frequency between inner and outer radii of the flow. Due to 
much longer viscous scales, the outer thin disc is not expected to be 
affected by global LT precession.  The emission of the precessing hot 
flow is then naturally modulated due a mixture of relativistic Doppler 
beaming, light bending and Compton anisotropy.  This model predicts the 
right range of observed LFQPO frequencies. The amplitude of the LFQPO 
depends on the details of the geometry and inclination of the viewing 
angle. The RMS is usually larger at high inclinations and can reach 10\,\% 
(see also \citealt{2015ApJ...807...53I} for predictions of 
modulation of the polarization of observed X-ray radiation).

In the truncated disc model the evolution of the disc inner radius drives 
both the evolution of the photon spectrum and the timing features. As the 
disc inner radius decreases, $\nu_{b}$ and the LFQPO frequencies are 
observed to move up, as expected in the model. At the same time, varying 
the truncation radius change the number of seed soft photons seen by the 
hot flow and the X-ray spectrum softens as the hot flow becomes gradually 
more efficiently cooled. Such softening of the spectrum is also observed 
\citep[see e.g.][]{2007A&ARv..15....1D}.
 

\begin{figure} 
\includegraphics[scale=0.25]{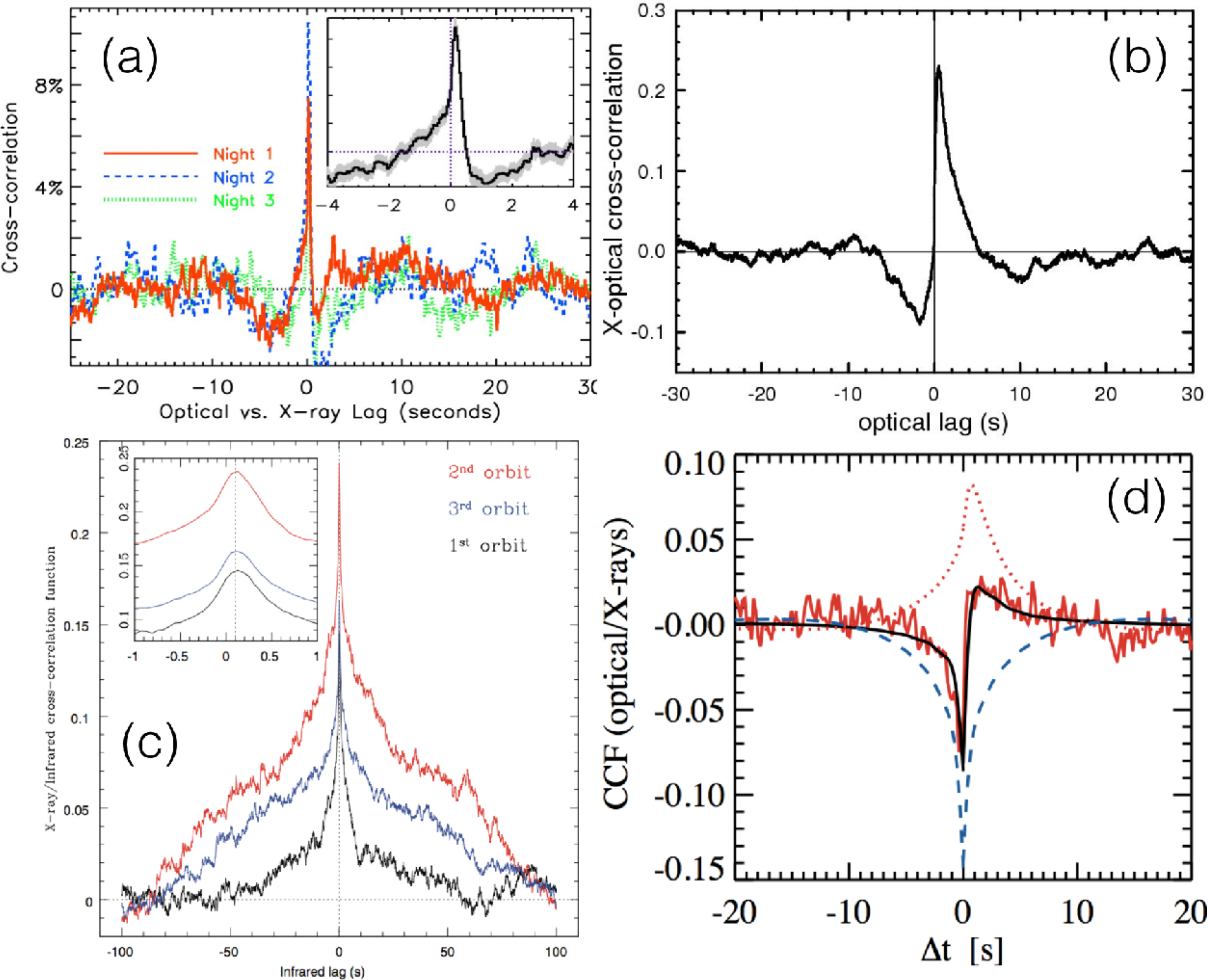} 
\caption{Various observed  OIR vs X-ray CCFs. Panel (a),(b) and (d) show
observed Optical vs X-ray CCFs of  GX\,339--4 
\citep{2008MNRAS.390L..29G}, XTE\,J1118+480 
\cite{2001Natur.414..180K} and Swift\,J1753-0127 \citep{2011ApJ...737L..17V} 
respectively. The CCF of Swift\,J1753-0127  is modelled with the reprocessing plus hot flow synchrotron model.
Panel (c) displays the IR / X-ray CCF observed in GX\,339--4 by 
\citet{2010MNRAS.404L..21C}. \copyright AAS. Reproduced with permission.
} 
\label{fig:ccfs} 
\end{figure}

\subsection{Optical timing}

Besides fast X-ray variability, several accreting BH sources show a  strong
variability in optical presenting comparable amplitude and PDS 
\citep[for e.g. see][]{2009ApJ...697L.167G,2010MNRAS.407.2166G}. 

\subsubsection{Reprocessing model}

Part of this variability could be associated to the reprocessing of the X-ray
illumination in the outer disc and at the surface of the companion star. If so,
the optical reprocessed flux $O(t)$ is expected to respond linearly to the X-ray
flux  $X(t)$:

\begin{equation}
O(t)\,=\,\int_{-\infty}^{t}X(t') r(t-t') dt'
\end{equation}

\noindent
The response function $r$ can be calculated for given geometry. Such 
calculations are presented in \citet{2002MNRAS.334..426O}. The 
reprocessing model predicts that the optical should be correlated with 
the X-ray flux and lagging behind the X-ray light curve by a few seconds 
(light travel time). Moreover the optical is expected to vary on longer 
time-scales than the X-rays, because the sub-second time-scale 
variability is strongly damped.  Reprocessing is a mechanism that is 
difficult to avoid and yet these predictions are not always verified.

\subsubsection{Fast optical variability from the accretion flow}

In several sources the optical fluctuations are observed on comparable or 
even faster time-scales than the X-rays. The cross-correlation function 
(CCF) of the X-ay and optical light curves show complex features involving 
also some level of anti-correlation (see Figure\,\ref{fig:ccfs}). The 
optical time lags can be as short as 100\,ms. This suggests an additional 
variability mechanism in optical. Such a mechanism could be the variable 
contribution of synchrotron emission by non-thermal coronal particles 
\citep{2011ApJ...737L..17V,2014SSRv..183...61P}. This emission could be 
almost as fast as the X-ray fluctuations. In the hybrid 
thermal-non-thermal hot flow model, moderate fluctuations of the mass 
accretion rate $\dot{m}$ lead to a simultaneous increase of the plasma 
Thomson depth and radiation power. Numerical simulations have shown this 
involves an anti-correlation between X-ray and optical fluxes, which is 
observed in several sources. \citet{2015MNRAS.448..939V} show that the 
X-ray and optical variability properties of the source Swift\,J1753-0127 
(shown in Figure\,\ref{fig:ccfs}) are well reproduced by a simple model 
assuming that the optical variability is a mixture of X-ray reprocessing 
plus a component that is exactly anti-correlated to the X-rays (as 
expected in the hot flow synchrotron model).

LF QPOs are observed in several hard state sources in the optical band 
with RMS amplitude ranging from 3 to 30\,\%. The optical QPOs are 
sometimes associated to the X-ray LF QPO but not always.  These QPOs 
could originate from the same LT precession mechanisms leading to the 
formation of the X-ray QPOs. The dominant contribution to the optical 
synchrotron flux comes from partially self-absorbed synchrotron. The 
effective synchrotron self-absorption depth depends on viewing angle. 
Therefore the specific synchrotron intensity produced by the hot flow is 
modulated by precession

\begin{equation}
I_s\propto 1-\exp\left[-\tau/\cos{\theta(t)}\right]
\end{equation}

\noindent
Depending on the geometric parameters this can lead to QPO amplitudes 
comparable to those observed \citep[see][]{2013MNRAS.430.3196V}. In 
principle it is also possible to predict the phase lags between the X-ray 
and optical QPOs. Assuming a simple prescription for the angular 
dependence of the hot flow Comptonized radiation, 
\citep{2013MNRAS.430.3196V} predict that the X-ray and optical QPO should 
be either in phase or in opposition (phase 0 or $\pi$).  
\citet{2015MNRAS.454.2855V} show that in Swift\,J1753.5-0127 the optical 
and X-ray QPOs are in phase, as predicted by the model. In this source at 
least the X-ray/optical CCF, the optical and X-ray PDS, their phase lags 
and coherence spectra are well accounted for by the synchrotron emitting 
hot flow model.

Another possible origin for the optical LFQPO could be disc reprocessing. 
Indeed the varying disc illumination caused by the LT precession of the 
hot flow naturally leads to the modulation of the reprocessed radiation 
(Veledina \& Poutanen 2015). In this case, a QPO is expected only if the 
hot flow precession period is longer than light crossing time of the disc 
which is itself a function of the orbital separation of the binary 
system

\begin{equation}
\nu_{QPOmax}\simeq \frac{c}{R_{disc}}\simeq  \frac{2}{3}\left(\frac{P_{\rm orb}}{1 {\rm hr}}\right)^{-2/3}\left(\frac{M}{10 M_{\odot}}\right)^{-1}  \qquad {\rm Hz}.
\end{equation}

\noindent
The amplitude of the QPO, its pulse profile and the expected optical vs X-ray
QPO phase lags are  expected to depend on the QPO frequency.  So far none of the
observed optical QPOs matches these properties and are more likely to be caused
by synchrotron.

\subsection{Caveats}

In the previous section, the current `standard' model for the emission of 
accretion flows around BHs was described. It involves a truncated disc 
with variable inner radius, a hot inner flow harbouring non-thermal 
electrons, fluctuations of the mass accretion rate propagating radially 
inward, and precession of the hot accretion flow.  This complex model 
produces the main features of the emission. In particular the observed 
simultaneous evolution of the SEDs and X-ray PDS can be understood as 
driven by changes in the inner disc radius. The observed optical 
variability and and the correlations between X-ray and optical bands are 
qualitatively understood.  There are however a few issues with this model 
that should be mentioned.  First, the detailed comparison of the hybrid 
magnetised model with observations has brought interesting constraints on 
the magnetic field. In the hard state in particular, the conclusion is 
that either the magnetic field is strongly sub-equipartition (which would 
be in contradiction with theoretical models involving a magnetically 
dominated accretion flow, or accretion disc corona atop the disc), or, 
the MeV excess is produced in a region that is spatially distinct from 
that producing the bulk of the hard X-ray radiation 
\citep{2013MNRAS.430..209D}.

In the latter case, the accretion flow could for instance, be constituted 
of a truncated accretion disc surrounding a central magnetically 
dominated hot accretion flow, responsible for the thermal Comptonisation 
component, while active coronal regions above and below the outer disc 
may produce the mostly non-thermal Compton emission (i.e. the hard state 
MeV excess, Malzac 2012). In the soft state the disc inner radius moves 
inwards until it reaches the ISCO and the hard X-ray emission becomes 
gradually dominated by the non-thermal corona. But if the thermal and 
non-thermal particles are not in the same location, the observed 
correlations between the X-ray variability and optical synchrotron 
variability cannot be explained by simultaneous changes in Thomson depth 
and luminosity. A different interpretation would be required.  Another 
interesting possibility is suggested by INTEGRAL measurements showing 
that the hard state MeV tail of Cyg\,X--1 is strongly polarised (at a level 
of 70 $\pm$30\,\%), while the thermal Comptonisation emission at 
lower energy is not \citep{2011Sci...332..438L,2012ApJ...744...64J}

The only plausible mechanisms for the very high level of polarisation 
seems to be synchrotron emission in a highly coherent magnetic field. A 
natural explanation would be that the MeV excess is in fact a 
contribution from the jet  \citep{2014A&A...562L...7R}.
Conventional jet models can produce such MeV component but require some 
fine tuning and quite extreme acceleration parameters 
\citep[see][]{2014MNRAS.442.3243Z}. If the non-thermal MeV excess is 
caused by jet synchrotron emission this strongly reduces the possible 
number of non-thermal electrons in the corona.  If so, the coronal optical 
synchrotron component would be drastically reduced and again, the fast 
optical variability becomes difficult to explain.  As will be shown in 
the Section\,\ref{sec:bhbjets}, 
a possible solution to these problems is to consider 
the contribution of the jets to the optical and IR emission.

\section{Model for the jets}
\label{sec:bhbjets}

\begin{figure} 
\includegraphics[width=5cm]{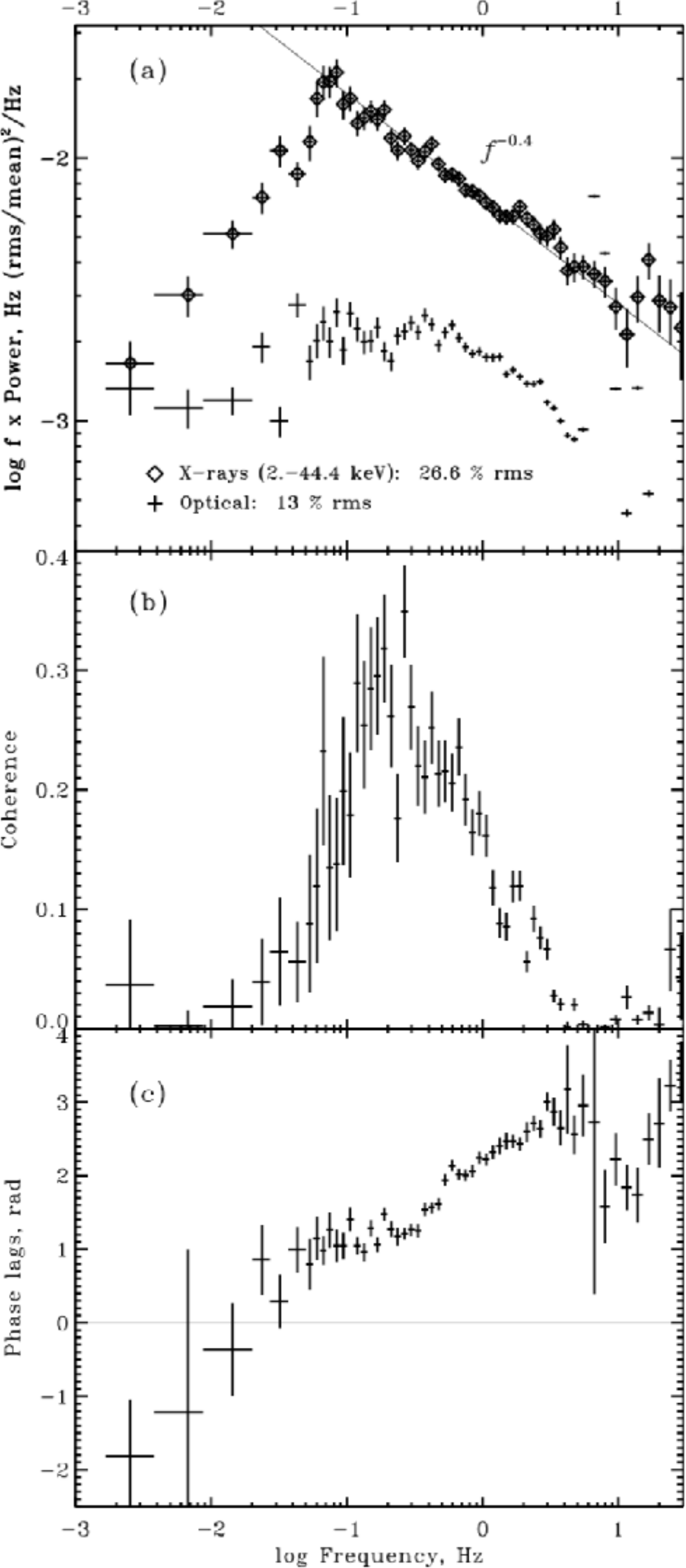} 
\caption{The optical/X-ray correlations of XTE\,J1118+480 in the Fourier domain. a) X-ray and optical power spectra. 
The counting noise was subtracted (see Section\,3.1). b) X-ray/optical coherence. c) 
phase-lags as function of Fourier frequency. A positive lag implies that the optical is delayed with respect 
to the X-rays \citep[][]{2003A&A...407..335M}. 
Reproduced with permission \copyright ESO.}
\label{fig:four1118} 
\end{figure}

In this section we will focus on the emission of BHB jets. The 
contribution of jets to the X-ray emission of BHBs is a controverted 
issue but there is evidence that in the most documented sources the jet 
X-ray emission is weak compared to the emission from the hot flow/corona 
(see discussion in Malzac 2015 and reference therein). Therefore in this 
section we will adopt the conventional view that the X-rays are dominated 
by the accretion flow.  Instead, we will focus on the contribution of 
jets to the observed fast variability in IR and optical bands (hereafter 
OIR). Indeed, in OIR, the combination of a reprocessed and a hot flow 
synchrotron component that is anti-correlated to the X-ray reproduces 
remarkably the shape of the optical/X-ray CCFs of sources like Swift\,J1753.5-0127 which show a strong opt/X-ray anti-correlation and a very 
weak reprocessing peak in their CCFs. However several sources such as GX\,339--4 or 
XTE\,J1118+480 have an optical X-ray CCF that is very difficult 
to model because the main peak of the correlation is too narrow to be due 
to reprocessing and the lag is too small (see Figure\,\ref{fig:ccfs} for a 
comparison). This suggests some ingredient is missing in the model. 
Moreover, timing observations of GX\,339--4 in IR also show similarly 
complicated CCFs with sometimes very sharp IR response lagging behind by 
approximatively 100\,ms. Even if the hot flow contains non-thermal 
electron their synchrotron emission is unlikely to be strong at IR 
wavelength.  The IR variability features have been naturally ascribed to 
the jets, which are indeed expected to produce IR synchrotron radiation. 
The 100\,ms lag was associated to the travel time for a fluctuation to 
travel from the disc to jet IR emitting region 
\citep[C10][]{2010MNRAS.404L..21C}. In the following, it will be shown 
that in fact, the jet could be responsible for most of the observed OIR 
variability. The interesting consequence is that the observed OIR/X-ray 
correlation may tell us something about the dynamics of accretion and 
ejection processes.  Such jet models for the OIR variability were 
initially developed in the early 2000 in the context of the 
multi-wavelength observations of a very exciting BHB which had recently 
been discovered: XTE\,J1118+480 \citep{2000IAUC.7389....2R}.  Its 
relatively close distance ($\simeq\,1.8$ kpc and the exceptionally low 
interstellar extinction towards the source \citep{2000IAUC.7392....2G} 
allowed a multi-wavelength monitoring of the source during its outburst 
(C03 and reference therein). During the whole outburst duration, the 
X-ray properties of the source, as well as the presence of strong radio 
emission, were typical of BH binaries in the hard state. In the radio to 
optical bands, a strong non-thermal component was associated with 
synchrotron emission from a powerful jet or outflow 
\citep{2001MNRAS.322L..23F}. Interestingly, fast optical and UV 
photometry allowed by the weak extinction, revealed a rapid optical/UV 
flickering presenting complex correlations with the X-ray variability 
(\citealt{2001Natur.414..180K,2003MNRAS.345..292H}, hereafter K01 and H03 
respectively).  This correlated variability cannot be caused by 
reprocessing of the X-rays in the external parts of the disc. Indeed, the 
optical flickering occurs on average on shorter time-scales than the 
X-rays (K01, see Figure\,\ref{fig:four1118}a), and reprocessing models 
fail to fit the complicated shape of the X-ray/optical cross correlation 
function (H03, see Figure\,\ref{fig:ccfs}b). Spectrally, the jet emission 
seems to extend at least up to the optical band 
(\citet{2001ApJ...555..477M}, hereafter C03), although the external parts 
of the disc may provide an important contribution to the observed flux at 
such wavelengths. The jet activity is thus the most likely explanation 
for the rapid observed optical flickering. For this reason, the 
properties of the optical/X-ray correlation in XTE\,J1118+480 might be of 
primary importance for the understanding of the jet-corona coupling and 
the ejection process.

The observations are very challenging for any accretion model. The most 
puzzling pieces of evidence are the following: (a) the optical/X-ray CCF 
shows the optical band lagging the X-ray by $\,$0.5 s, but with a dip 2-5 
seconds in advance of the X-rays (K01); (b) the correlation between X-ray 
and optical light curves appears to have time-scale invariant properties: 
the X-ray/optical CCF maintains a similar, but rescaled, shape on 
time-scales ranging at least from 0.1 s to few tens of seconds 
\citep[][hereafter M03]{2003A&A...407..335M}; (c) the correlation does 
not appear to be triggered by a single type of event (dip or flare) in 
the light curves; instead, as was shown by M03, optical and X-ray 
fluctuations of very different shapes, amplitudes and time-scales are 
correlated in a similar way, such that the optical light curve is related 
to the time derivative of the X-ray one. Indeed, in the range of 
time-scales where the coherence is maximum, the optical/X-ray phase lag 
are close to $\pi/2$ (see Figure\,\ref{fig:four1118}). This indicates the 
two light curves are related trough a differential relation. Namely, if 
the optical variability is representative of fluctuations in the jet 
power output $P_{\rm j}$, the data suggest that the jet power scales 
roughly like $P_{\rm j} \propto -\frac{dP_{\rm x}}{dt}$, where $P_{\rm 
x}$ is the X-ray power.

\subsection{The energy reservoir model}

\citet[][hereafter MMF04]{2004MNRAS.351..253M} have shown that the 
complex X-ray/optical correlations could be understood in terms of an 
energy reservoir model.  In this picture, it is assumed that large 
amounts of accretion power are stored in the accretion flow before being 
channelled either into the jet (responsible for the variable optical 
emission) or into particle acceleration/ heating in the Comptonizing 
region responsible for the X-rays.  MMF04 have developed a time dependent 
model which is complicated in operation and behaviour. However, its 
essence can be understood using a simple analogue: Consider a tall water 
tank with an input pipe and two output pipes, one of which is much 
smaller than the other. The larger output pipe has a tap on it. The flow 
in the input pipe represents the power injected in the reservoir $P_{\rm 
i}$, that in the small output pipe the X-ray power $P_{\rm x}$ and in the 
large output pipe the jet power $P_{\rm j}$.  If the system is left alone 
the water level rises until the pressure causes $P_{\rm i}\,=\,P_{\rm 
j}+P_{\rm x}$.  Now consider what happens when the tap is opened more, 
causing $P_{\rm j}$ to rise. The water level and pressure (proportional 
to $E$) drop causing $P_{\rm x}$ to reduce. If the tap is then partly 
closed, the water level rises, $P_{\rm j}$ decreases and $P_{\rm x}$ 
increases. The rate $P_{\rm x}$ depends upon the past history, or 
integral of $P_{\rm j}$. Identifying the optical flux as a marker of 
$P_{\rm j}$ and the X-ray flux as a marker of $P_{\rm x}$ we obtain the 
basic behaviour seen in XTE\,J1118+480.  In the real situation, MMF04 
envisage that the variations in the tap are stochastically controlled by 
a shot noise process. There are also stochastically-controlled taps on 
the input and other output pipes as well. The overall behaviour is 
therefore complex. The model shows however that the observed complex 
behaviour of XTE\,J1118+480 can be explained by a relatively simple basic 
model involving several energy flows and an energy reservoir.  This 
simple model is largely independent of the physical nature of the energy 
reservoir. In a real accretion flow, the reservoir could take the form of 
either electromagnetic energy stored in the X-ray emitting region, or 
thermal (hot protons) or turbulent motions. The material in the disc 
could also constitute a reservoir of gravitational or rotational energy 
behaving as described above. \label{sec:model} In a stationary flow, the 
extracted power $P_{\rm j}+P_{\rm x}$ would be perfectly balanced by the 
power injected, which is, in the most general case, given by the 
difference between the accretion power and the power advected into the 
hole and/or stored in convective motions: $P_i\,\simeq\,\dot M c^2 - P_{\rm 
adv,conv}$. However, observations of strong variability on short time 
scale clearly indicate that the heating and cooling of the X-ray (and 
optical) emitting plasma are highly transient phenomena, and the corona 
is unlikely to be in complete energy balance on short time-scales. MMF04 
therefore introduced a time-dependent equation governing the evolution of 
its total energy $E$

\begin{equation}
\dot E\,=\, P_{\rm i} - P_{\rm j} - P_{\rm x},
\label{eq:enbal}
\end{equation}

\noindent
and we assume that all the three terms on the right hand side are time
dependent. The optical variability is produced mainly from  synchrotron emission
in the inner part of the jet at distances of a few thousands gravitational radii
from the hole. We assume that  at any time the optical flux $O_{pt}$ (resp.
X-ray flux) scales like the jet power $P_{\rm j}$ ( plasma heating power $P_{\rm
x}$). MMF04 introduced the instantaneous dissipation rates $K_{\rm j}$ and
$K_{\rm x}$

\begin{equation}
 P_{\rm j}(t)\,=\, K_{\rm j}(t)E(t),  
 \qquad P_{\rm x}(t)\,=\, K_{\rm x}(t)E(t), 
\label{def:kx}
\end{equation}

For a specific set of parameters MMF04 generate random independent fluctuations
(time series) for $K_{\rm x}$, $K_{\rm j}$ and $P_{\rm i}$, solve the time
evolution of the energy reservoir $E$ and then use the solution to derive the
the resulting optical and X-ray light curves (see MMF04 for details).

\begin{figure*}
\centerline{\scalebox{0.4}{\includegraphics{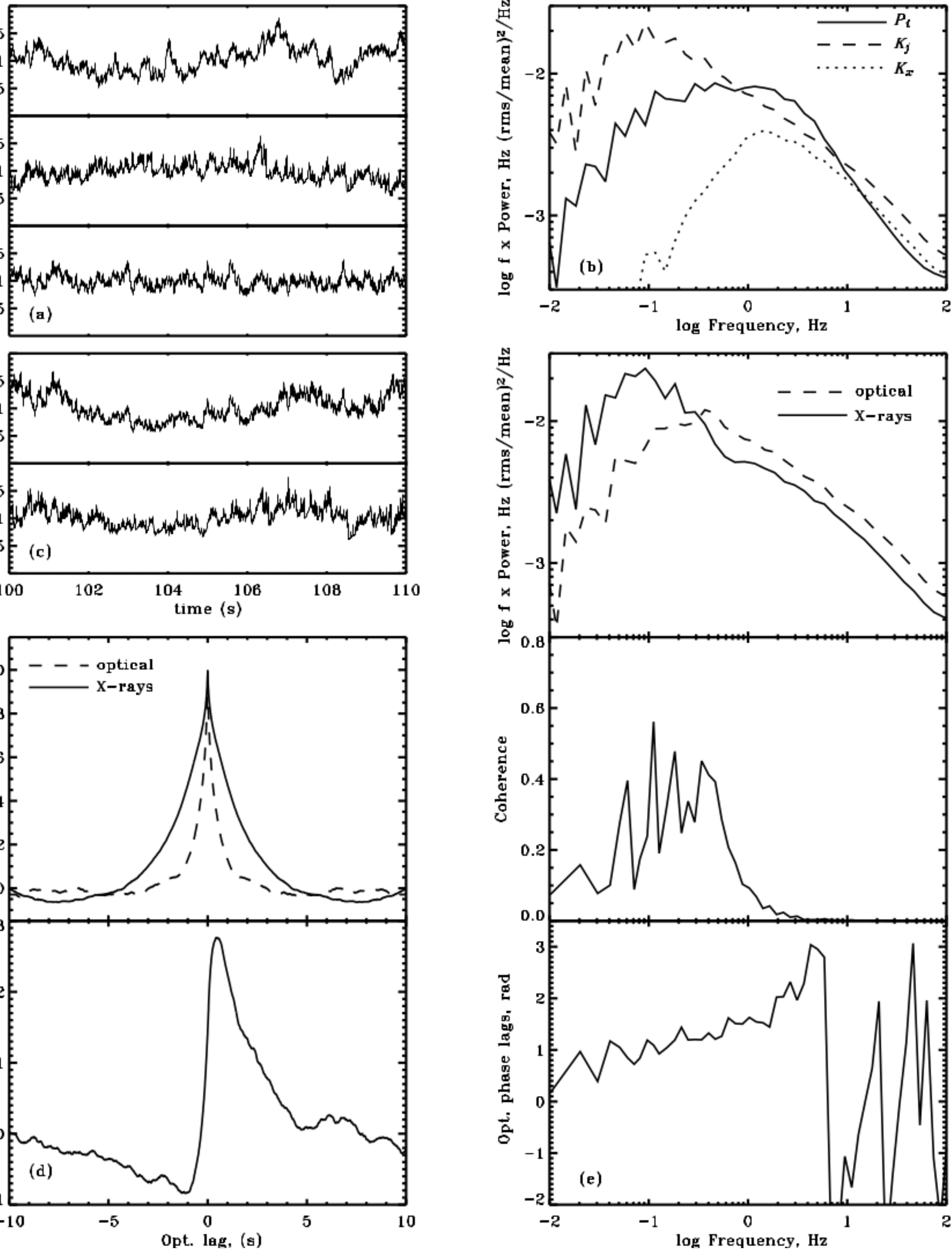}}}
\caption{Sample input time series (panel a) and power
spectra (panel b) of $P_{\rm i}$,
$K_{\rm j}$, $K_{\rm x}$, resulting X-ray and optical fluxes light curves
(panel c),  X-ray/optical autocorrelation and cross-correlation
functions (panel d), power spectra, coherence and phase-lags (panel e)
\citep[From][]{2004MNRAS.351..253M}.} 
\label{fig:simu184}
\end{figure*}

Combining equations (\ref{eq:enbal}) and (\ref{def:kx}) we obtain the
following relation for the total instantaneous jet power

\begin{equation}
P_{\rm j}\,=\,P_{\rm i} - (1 + \frac{\dot K_{\rm x}}{K_{\rm x}^2})P_{\rm x} -
\dot P_{\rm x}/K_{\rm x}.
\label{eq:jdom}
\end{equation}

We can see from this equation that the differential scaling $P_{\rm j} 
\propto - \dot P_{\rm x}$, observed in XTE\,J1118+480, will be rigorously 
reproduced provided that:
(1) $K_{\rm x}$ is a constant;
(2) $P_{\rm i} - P_{\rm x}$ is a constant.
It is physically unlikely that those conditions will be exactly verified. 
In particular, $P_{\rm x}$ is observed to have a large RMS amplitude of 
variability of about 30\,\%. However, the observed differential 
relation holds only roughly and only for fluctuations within a relatively 
narrow range of time-scales $1-10 s$. Therefore, the above conditions 
need only to be fulfilled approximatively and for low frequency 
fluctuations ($> 1$s). In practice, the following requirements will be 
enough to make sure that the low frequency fluctuations of the right hand 
side of equation $\ref{eq:jdom}$ are dominated by $\dot P_{\rm x}$:

\begin{itemize} 
\item $P_{\rm x} \ll P_{\rm i}$, implying that the jet power, on average, dominates over the X-ray luminosity;
\item the amplitude of variability of $K_{\rm x}$ and $P_{\rm i}$ in the
1--10\,s range is low compared to that of $P_{\rm j}$. In other words the
1--10\,s fluctuations of the system are mainly driven by the jet activity, 
implying that the mechanisms for dissipation in the jet and the corona occur
 on quite different time-scales.  
\end{itemize}

Figure\,\ref{fig:simu184} shows the results of a simulation matching the 
main timing properties of XTE\,J1118+480. In this simulation jet power was 
set to be 10 times larger than the X-ray power. As can be seen, the 
resulting PSDs, CCFs, phase-lag and coherence spectra are very similar to 
those observed in XTE\,J1118+480.

\begin{figure} 
\includegraphics[width=0.7\textwidth]{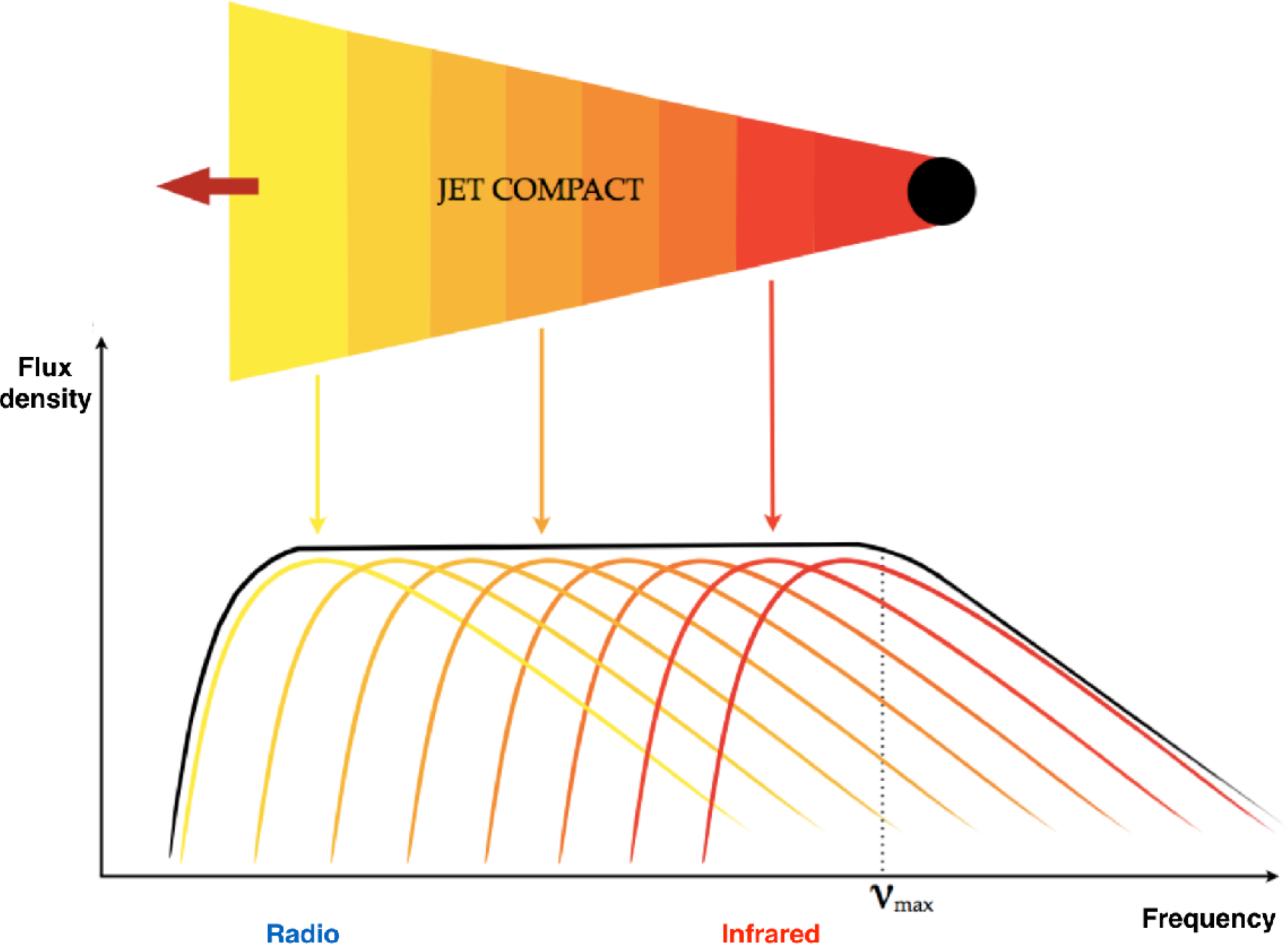} 
\caption{The jet SED is made of the contribution of 
synchrotron emitting regions distributed over the jet length 
\citep[][]{CoriatPHD}.} 
\label{fig:corjet} 
\end{figure}

\subsection{Internal shock model}

The puzzling optical/X-ray correlations of  XTE\,J1118+480, can be understood in
terms of a common energy reservoir for both the jet and the Comptonizing
electrons. However the model outlined above remains a toy model and more
realistic jet physics must be taken into account. In particular we have to
explain how jet radiation is produced.  The  flat SED of compact jets are
usually ascribed to self-absorbed synchrotron emission in a  conical jet
geometry \citep{1979ApJ...232...34B}. The model postulates the presence of a
standing  shock located close to the base of the jet (at distances ranging
between 10$^3$ to 10$^4$ $R_{\rm G}$ from the BH.  In this shock electrons are
accelerated with a power-law energy distribution extending up to
ultra-relativistic energies and then propagate along the jet with the flow while
radiating synchrotron radiation. As the electrons travel in this conical
geometry their density decreases and the peak of synchrotron emission (which
corresponds to the turn-over frequency between optically thin and optically
thick synchrotron, see Section\,\ref{sec:synchabs}) moves toward  longer
wavelengths.  The jet SED is made of the sum of the emitting electrons
distributed all along the the jet. The radio emission is produced at large
distance while the IR is produced close to the initial acceleration region (see
Figure\,\ref{fig:corjet}). A flat spectrum is produced under the assumption of
continuous energy replenishment of the adiabatic expansion losses.   Adiabatic
expansion losses are simply the internal energy losses of the gas (and tangled
magnetic field) due to pressure work against the external medium as it expands
in the conical geometry. The compensation of these energy losses is crucial for
maintaining this specific spectral shape \citep{2006MNRAS.367.1083K}. Some dissipation
mechanism is required to compensate for these losses otherwise the radio
emission is strongly suppressed and the jet SED is too inverted.

Internal shocks provide a possible mechanism to compensate for the adiabatic
expansion losses by dissipating energy and accelerating particles at large
distance from the BH. Internal shocks caused by fluctuations of the outflow
velocity are indeed widely believed to power the multi-wavelength emission of
jetted sources such as $\gamma$-ray bursts 
\citep{1994ApJ...430L..93R,1998MNRAS.296..275D},  active galactic nuclei 
\citep{1978MNRAS.184P..61R, 2001MNRAS.325.1559S}, or
microquasars \citep{2000A&A...356..975K,2010MNRAS.401..394J}. . Internal
shocks models usually assume that the jet can be discretised into homogeneous
ejectas. Those ejectas are injected at the base of the jet with variable
velocities and then propagate along the jet. At some point, the fastest
fluctuations start  catching up and merging with slower ones. This leads to
shocks in which a fraction of the bulk kinetic velocity of the shells is
converted into internal energy. Part of the dissipated energy goes into 
particles acceleration, leading to synchrotron and also, possibly, inverse
Compton emission. 

\begin{figure*}
 \includegraphics[scale=0.27]{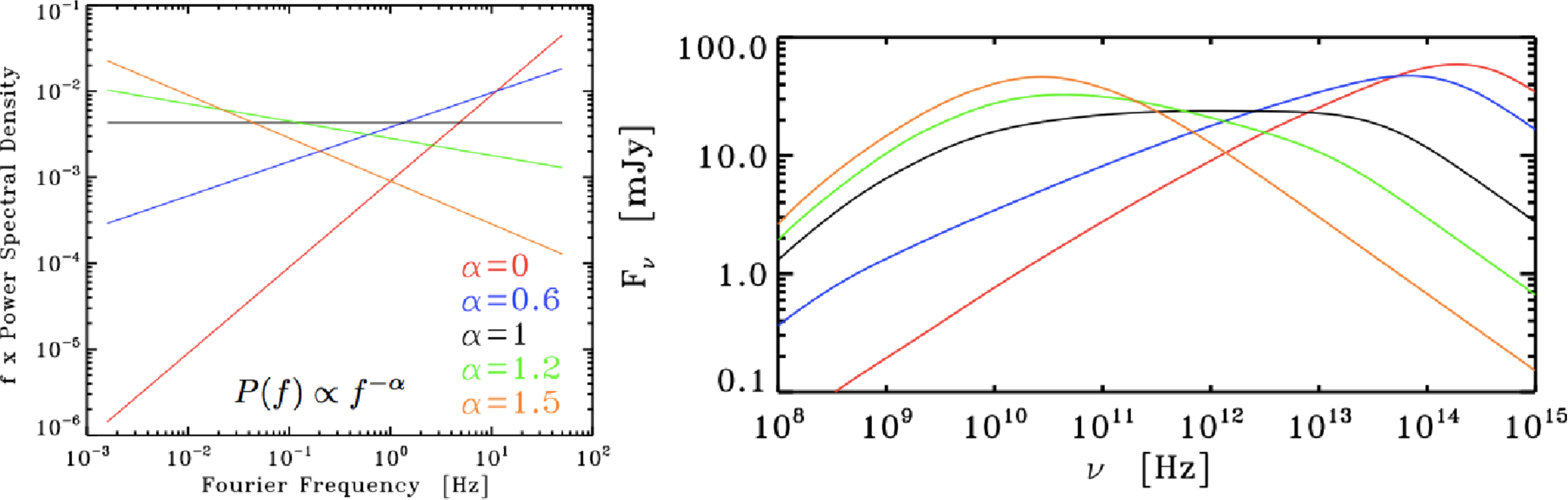}
 \caption{Simulation of the internal shock model with a power-law PSD of the
Lorentz factor fluctuations ($P(f)\propto f^{-\alpha}$). The top left panel
shows the  shape of the injected PSDs, for the indicated values of the $\alpha$
index. The right panel shows the jet SED  calculated for an inclination angle of
40 degrees and a distance to the source of 2 Kpc. \citep[see][for
details]{2014MNRAS.443..299M, 2015sf2a.conf..161M}}
 \label{fig:8}
 \end{figure*}

The energy dissipation profile of the internal shocks is very sensitive 
to the shape of the PSD of the velocity fluctuations.  Indeed, let us 
consider a fluctuation of the jet velocity of amplitude $\Delta v$ 
occuring on a time-scale $\Delta t$. This leads to the formation of a 
shock at a downstream distance $z_s \propto \Delta t /\Delta v$. In this 
shock the fraction of the kinetic energy converted into internal energy 
will be larger for larger $\Delta v$. From these simple considerations we 
see that the distribution of the velocity fluctuation amplitudes over 
their time-scales (i.e. the PSD) is going to determine where and in which 
amount the energy of the internal shocks is deposited.  
\citet{2012IJMPS...8...73M} used Monte-Carlo simulations to study this 
dependance and found that independently of the details of the model flat 
radio-IR SEDs are obtained for a flicker noise PSD of the fluctuations of 
the jet Lorentz factor.  This result is illustrated by 
Figure\,\ref{fig:8}, which compares the SEDs obtained for PSD of the 
Lorentz factor of the jet with a power-law shape with varying index 
$\alpha$: $P(f)\propto f^{-\alpha}$.  For larger $\alpha$ the 
fluctuations of the Lorentz factor have, on average, longer time-scales 
and therefore more dissipation occurs at larger distances from the BH. 
One can see from Figure\,\ref{fig:8} that the SED is very sensitive to the 
value of $\alpha$, for $\alpha\,=\,1$ (i.e. flicker noise) the dissipation 
profile scales like $z^{-1}$ and the specific energy profile is flat. In 
other words, the internal shocks compensate exactly for the adiabatic 
losses. As result the SED is flat over a wide range of photon 
frequencies. In fact, this result can also be obtained analytically 
(Malzac 2013). The case of flicker noise fluctuations of the jet Lorentz 
factor may therefore be relevant to the observations of compact jets.

\begin{figure*}
 \includegraphics[scale=0.26]{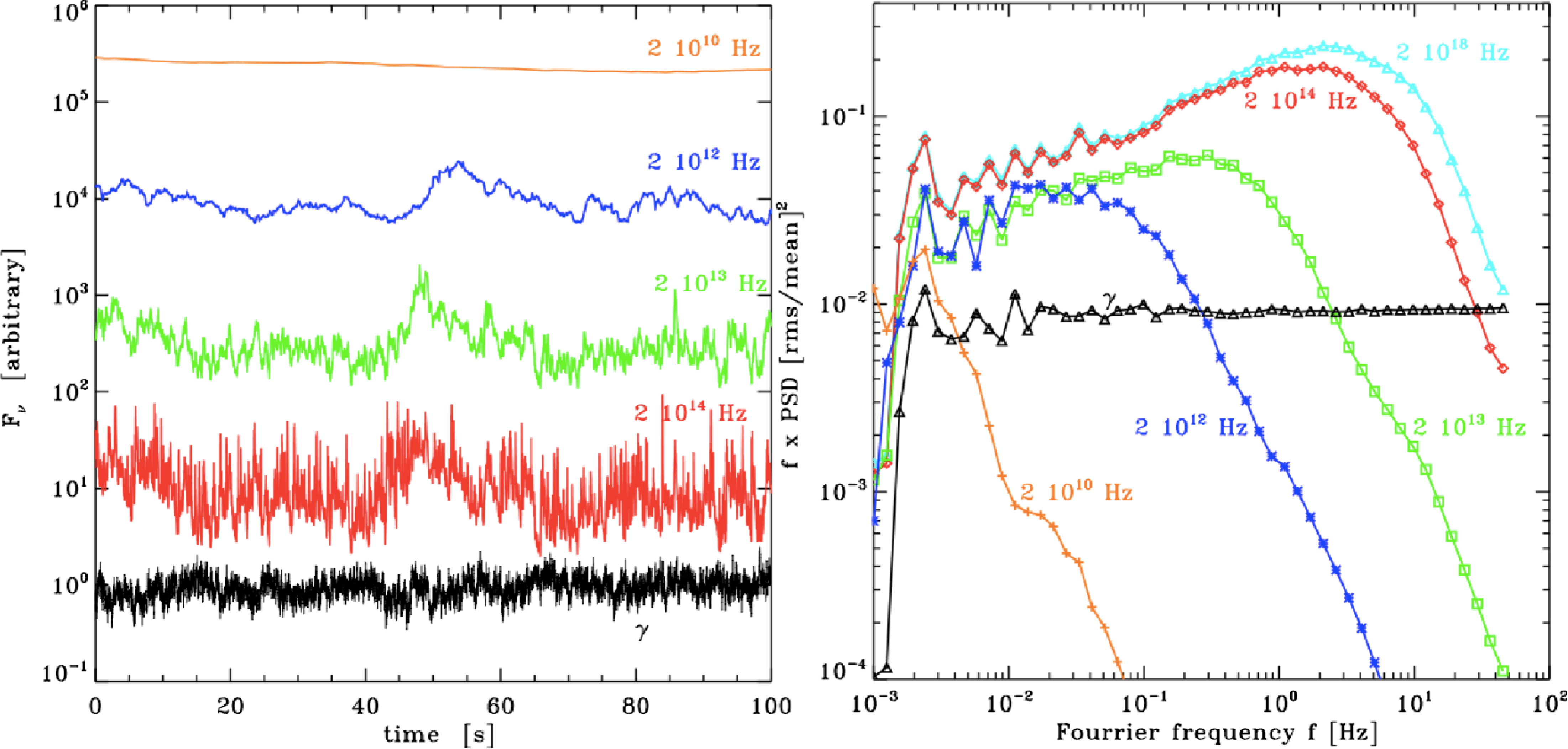}
 \caption{Synthetic light curves (left, rescaled) and power spectra at various 
indicated frequencies resulting from the simulation with $\alpha=0$. The
injected fluctuations of the Lorentz factor are also shown. 
 \citep[][]{2014MNRAS.443..299M}.}
 \label{fig:2}
 \end{figure*}

An interesting feature of the internal shock model is that it naturally 
predicts strong variability of the jet emission. Figure\,\ref{fig:2} shows 
sample light curves and power spectra obtained from the simulation with 
$\alpha\,=\,1$. The jet behaves like a low-pass filter. As the shells of 
plasma travel down the jet, colliding and merging with each other, the 
highest frequency velocity fluctuations are gradually damped and the size 
of the emitting region increases. The jet is strongly variable in the 
optical and IR bands originating primarily from the base of the emitting 
region and becomes less and less variable at longer frequencies produced 
at larger distances from the BH.

Note that strong coherent periodic fluctuations of of the jet Lorentz 
factor may lead to the formation of QPO features in the jet variability.  
However, in the context of the jet model a much more likely explanation 
for the observed optical and IR QPOs could be the precession of the jets. 
Indeed if the X-ray LF QPOs are caused by global LT precession of the hot 
flow and if the jet is launched from the accretion flow, one may expects 
the jet to precesses with the flow. The (mostly) optically thin 
synchrotron radiation observed in IR and optical would then be modulated 
at the precession frequency due to the modulation of Doppler beaming 
effects toward the observer \citep{2016MNRAS.460.3284K}.

The model may be used to understand the observed complex timing 
correlation between the X-ray and IR bands. For example, C10 measured de 
CCF of the X-ray and IR light curves and found significant correlation 
between the two bands with the IR photons lagging behind the X-rays by 
about 100\,ms (see Figure\,\ref{fig:ccfs}c). C10 interpreted this time-lag 
as the propagation time of the ejected shells from the accretion flow to 
the IR emitting region in the jet.  In the framework of the internal 
shock model this observation suggests that the fluctuations of the jet 
Lorentz factor are related to the X-ray variability of the source.  
Malzac , 2014 shows that the internal shock model predict very similar IR 
vs X-ray CCF and lags. provided that the fluctuation of the bulk jet 
Lorentz factor are inversely proportional to the X-ray flux.

\subsection{Are the jet Lorentz factor fluctuations related to the X-ray
variability ?} 

In fact, if the jet is launched from the accretion disc, the variability 
of the jet Lorentz factor must be related to that of the accretion disc. 
And we know, both from theory \citep[see e.g.][]{1997MNRAS.292..679L} and 
from observations\citep[see e.g.][]{2005astro.ph..1215G} that accretion 
discs tend to generate flicker noise variability.  Therefore flicker 
noise fluctuations of the jet Lorentz factor are not unexpected. In X-ray 
binaries the variability of the accretion flow is best traced directly by 
the X-ray light curves. The left panel of Figure\,\ref{fig:1bis} shows an 
actual X-ray PDS oserve in GX\,339--4 in the hard state, which differs 
significantly from flicker noise both at high and low frequencies. This 
variability could be a good proxy for the assumed fluctuations of the 
jet. The right panel of Figure\,\ref{fig:1bis} shows the resulting time 
averaged SED obtained if one assume that the Lorentz factor fluctuations 
have a PSD that is exactly that observed in X-rays. This synthetic SED is 
compared to multi wavelength observation that are nearly simultaneous 
with the X-ray timing data \citep[see][for details]{2015MNRAS.447.3832D}. 
The model appears to reproduce pretty well the radio to IR data.  This 
agreement is striking because the shape of the SED depends almost 
uniquely on the assumed shape of the PSD of the fluctuations. Although 
the model has a number of free parameters (jet power, inclination angle, 
time-averaged jet Lorentz factor...) that could be tuned to fit the data, 
those parameters only affect the flux normalisation or shift in the 
photon frequency direction, but they have very little effects on the 
overall shape of the SED.

\begin{figure*}
   \includegraphics[scale=0.26]{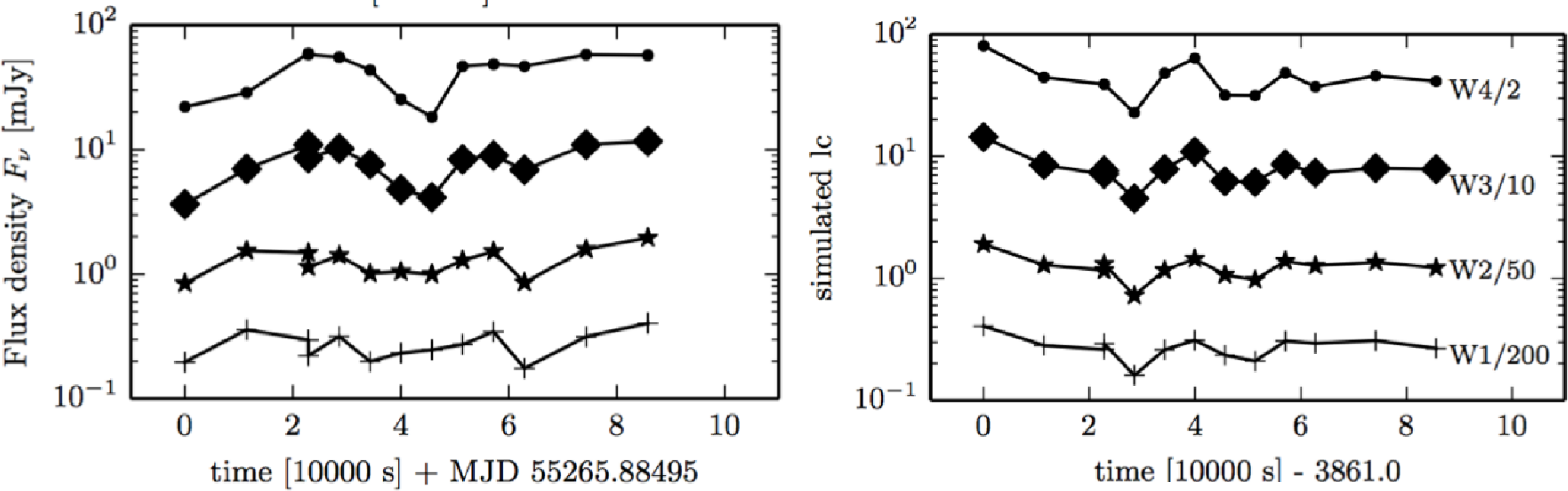}
 \caption{Left: the observed mid-IR variability as observed by WISE in 4 bands. 
 Right: sample synthetic light curves obtained from the same simulation shown 
 in Figure\,\ref{fig:1bis}. 
 \citep[][]{2015sf2a.conf..161M}.
}
 \label{fig:2bis}
 \end{figure*}

The four mid-IR flux measurements at 1.36$\times10^{13}$, 
2.50$\times10^{13}$, 6.52$\times10^{13}$ and 8.82$\times10^{13}$\,Hz that 
are shown on Figure\,\ref{fig:1bis} were obtained with the Wide field 
Infrared Survey Explorer \citep[WISE][]{2010AJ....140.1868W} they 
represent an average over 13 epochs, sampled at multiples of the 
satellite orbital period of 95 minutes and with a shortest sampling 
interval of 11 s, when WISE caught the source on two consecutive scans. 
These data have revealed a strong variability of the mid-IR emission (see 
G11). The light curves of these observations are shown in 
Figure\,\ref{fig:2bis} (left panel) and compared to light curves obtained 
from the same simulation that gives a good fit to the observed radio-IR 
SED (right panel).  The model appears to predict a variability of similar 
amplitude to that observed by WISE.


\section{Conclusion}

Compact objects can release huge amounts of energy on very short 
time-scales through extraction of their own rotational energy, 
dissipation of their powerfull magnetic fields, of the gravitational 
energy of accreted material which can be also gravitationally compressed 
to the point of thermonuclear fusion. A fraction of this energy is 
converted into radiation that is emitted non-thermally through 
synchrotron, curvature radiation inverse Compton, Bremsstrahlung. Based 
on the properties of these radiation processes and the energy dissipation 
mechanisms, time-dependent emission models are developed which, through 
comparison to data, allow to constrain the physical conditions in the 
emitting regions. Such comparisons may also may shed new lights on the 
physics of compact objects requiring in turn to revise the models.  In 
the the particular case of accreting BHBs, the truncated disc model 
including a hot inner flow with non-thermal electrons, fluctuations of 
the mass accretion rate propagating radially inward, and precession of 
the hot accretion flow produces the main features of the multi-wavelength 
spectra and variability. In particular in some sources, the observed 
optical variability and and the correlations between X-ray and optical 
bands are qualitatively understood.  However in some other sources a 
strong contribution of the jet to the IR and perhaps even optical 
variability seem to be required.  Fast jet synchrotron variability may 
results from the disc variability through a tight coupling between jet 
and accretion flow on short time-scales, possibly through a common energy 
reservoir. The IR jet radiation is likely to produced through synchrotron 
emission of particules accelerated in colliding shells of gas (internal 
shocks). This internal shock model naturally predicts the formation of 
the observed SEDs of compact jets and also predict a strong, wavelength 
dependent, variability that resembles the observed one. The model also 
suggests a strong connection between the observable properties of the jet 
in the radio to IR bands, and the variability of the accretion flow as 
observed in X-rays. If the model is correct, this offers a unique 
possibility to probe the dynamics of the coupled accretion and ejection 
processes leading to the formation of compact jets. Future multifrequency 
HTRA observations combined with further modelling may reveal this fast 
dynamic coupling.

\section{Acknowledgments} 

I am extremely grateful to the organisers of this Winter School, T. 
Shahbaz, J. Casares and T. Mu\~{n}oz Darias, for inviting me at the 
Instituto de Astrof\'isica de Canarias and for the opportunity to deliver 
these lectures. I also want to thank my long-time collaborator Renaud 
Belmont for his inputs, in particular the material he provided for the 
section on radiation processes. The preparation of this manuscript was 
supported in part by the ANR CHAOS project ANR-12-BS05-0009 
(http://www.chaos-project.fr).

\bibliography{Malzac/malzac}
\bibliographystyle{cambridgeauthordate}